\documentclass[a4paper,11pt]{article}
\pdfoutput=1 
\usepackage{jinstpub} 
\usepackage{epsfig}
\usepackage{bm}
\usepackage[pdftex]{color}
\title{\boldmath Implementation of Maxwell's equations in the
reconstruction of the magnetic field in the $g-2$ storage ring}

\author[a]{Geoffrey~T.~Bodwin,}
\author[a,b,c,1]{Hee~Sok~Chung \note{Corresponding author.}}
\author[a]{and Jos\'e Repond}
\affiliation[a]{High Energy Physics Division, Argonne National Laboratory,\\
9700 S. Cass Ave, Lemont, IL 60439, USA}
\affiliation[b]{Physik-Department, Technische Universit\"at M\"unchen,\\
James-Franck-Str. 1, 85748 Garching, Germany}
\affiliation[c]{Theory Department, CERN,\\ 1211 Geneva 23, Switzerland}

\emailAdd{heesok.chung@tum.de}

\abstract{
We present a method for implementing the constraints that are
implied by Maxwell's equations in fits to measurements of the magnetic
field in the muon storage ring of the $g-2$ experiment. The method that
we use makes use of toroidal-harmonic solutions of Laplace's equation.
We point out that the fitting problem can be approximated well as a
linear-algebra problem. We have devised an efficient algorithm for
the linear-algebra problem that makes it possible to find a solution
for $10^5$ data points and $10^4$ harmonics in less than an hour on a
present-day desktop computer. We illustrate our method by applying it to
some preliminary measurements of the magnetic field in the $g-2$ storage
ring.
}
\arxivnumber{1805.12163}
\begin{document}
\maketitle
\flushbottom

\section{Introduction}
\label{sec:intro}
Precise knowledge of the magnetic field inside the muon storage ring is
of paramount importance for the measurements of the muon anomalous
magnetic moment ($g-2$) that will be made by the Fermilab experiment
E989 \cite{tech-design-report}. The primary method of measurement of the
magnetic field, which utilizes a circular array of
nuclear-magnetic-resonance (NMR) probes, determines only the magnitude
of the magnetic field, not its direction. Ancillary measurements using
Hall probes have been used to obtain information about components of the
field that are orthogonal to its average direction \cite{Hall-probe-data}.

Additional information about the magnetic field can be obtained
by making use of the constraints that are implied by Maxwell's equations.
Since the inside of the muon storage ring is a vacuum, 
Maxwell's equations for the magnetic field $\bm{B}$ become
\begin{subequations}\label{maxwell-eqns}%
\begin{eqnarray}
\label{eq:maxwell1}%
\bm{\nabla} \cdot \bm{B} &=& \bm{0},\label{divergence} 
\\
\label{eq:maxwell2}%
\bm{\nabla} \times \bm{B} &=& \bm{0}.\label{curl} 
\end{eqnarray}
\end{subequations}
Because the magnetic field in the ring is almost entirely in the 
$z$ (vertical) 
direction, the NMR-probe measurements give information about the 
gradients of $B_z$. That information, in combination with 
eqs.~(\ref{maxwell-eqns}), allows one to compute the gradients of the field 
components in the transverse directions, which can be used to check
the Hall-probe measurements. In the $g-2$ experiment, the main effect 
of the transverse 
fields is to induce oscillations of the beam in the transverse 
direction, and so knowledge of the transverse fields provides a useful 
consistency test of the observations of beam dynamics. In a proposed 
measurement of the muon's electric dipole moment (EDM), knowledge of 
the transverse fields may play a more significant role, facilitating a 
reduction of measurement systematic uncertainties.

As a first step in analyzing the ring magnetic field, we make use of
the fact that the magnetic field is almost uniform in direction and
magnitude inside the storage ring. Therefore, we can write $\bm{B}$
as a sum of the uniform field $\langle B \rangle \hat{\bm{z}}$ and a
fluctuation $\bm{B}_{\rm f}$:
\begin{equation}
\label{eq:bfield_pert}%
\bm{B} = \langle B \rangle \hat{\bm{z}} + \bm{B}_{\rm f}, 
\end{equation}
where $\hat{\bm{z}}$ is a unit vector in the direction of the average of
the $\bm{B}$ field over the active volume of the storage ring. For the
E989 ring, $\hat{\bm{z}}$ is very close to the vertical direction. We
choose $\langle B \rangle$ to be about the same size as the average
magnetic field strength. The precise value of $\langle B \rangle$ is
unimportant, as small changes in $\langle B \rangle$ can be absorbed
into $\bm{B}_{\rm f}$. However, as we will see, it is crucial to our
analysis that $\bm{B}_{\rm f}$ be small compared with $\langle B
\rangle$.

Since the uniform field $\langle B \rangle \hat{\bm{z}}$ satisfies 
Maxwell's equations in vacuum, so does the fluctuation $\bm{B}_{\rm f}$. 
The equation $\bm{\nabla} \times \bm{B}_{\rm f} = \bm{0}$ 
implies that $\bm{B}_{\rm f}$ can be written as a gradient of a 
scalar function $V$, which satisfies 
\begin{subequations}
\begin{eqnarray}
\bm{\nabla} V &=& \bm{B}_{\rm f}, 
\\
\label{eq:laplace}%
\nabla^2 V &=& 0.
\end{eqnarray}
\end{subequations}
Eq.~\eqref{eq:laplace}, which follows from 
$\bm{\nabla} \cdot \bm{B}_{\rm f}=0$, is Laplace's equation.

The region of validity of eq.~\eqref{eq:laplace} can be an issue because
the single-valuedness of $V$ relies on eq.~(\ref{curl}), through Stokes'
theorem. For the E989 muon storage ring, there are no currents through
the plane of the ring at radii that lie inside the outermost part of the
vacuum chamber. Consequently, $V$ is single-valued inside the vacuum
chamber. 

In this paper, we  implement Maxwell's equations in fits to the NMR
measurements of the storage-ring magnetic field by expressing the fitting
function as a sum of solutions of Maxwell's equations. We do this by
writing $V$ as an expansion in a complete set of solutions of Laplace's
equation (harmonic functions). In particular, we use toroidal harmonics
as the basis functions, since their coordinate geometry is particularly
well matched to the geometry of a toroidal storage ring. We exclude
the large constant field $\langle B \rangle \hat{\bm{z}}$ from the
toroidal-harmonic fit, as it would be necessary to go to a very high
order in the toroidal harmonics in order to achieve a fit to $\langle B
\rangle \hat{\bm{z}}$ of sufficient accuracy.

The fitting problem for the $g-2$ ring involves a large number of data
points and harmonics, and, as we will see, is underconstrained. Hence,
the fitting problem, in its original form, is quite formidable.
We have developed several
simplifications/reorganizations of the fitting algorithm that make the
fitting problem computationally tractable.

We have applied our fitting algorithm to preliminary data from the
$g-2$ trolley runs. While these data do not reflect the ultimate
precision of the NMR measurements or the ultimate quality of the $g-2$
storage-ring magnetic field, they serve to illustrate the fitting
procedure and its utility.

The remainder of this paper is organized as follows. In
section~\ref{sec:toroidal} we describe the toroidal-harmonic expansion of
the magnetic field. In section~\ref{sec:fit} we outline our method for
fitting the toroidal harmonic series to the data. Here we describe the
algorithmic improvements that we have made. Section~\ref{sec:results}
contains some examples of the application of our fitting procedure to
some of the early NMR magnetic-field data. Finally, in
section~\ref{sec:summary}, we summarize our results.

\section{Use of toroidal harmonics to implement Maxwell's equations}
\label{sec:toroidal}

Toroidal-harmonic solutions of Laplace's equation of been known for many
years \cite{hicks-1881,basset-1897}. Many of the modern applications
of expansions of magnetic fields in toroidal harmonics have focused on
magnetic fields in tokamaks \cite{milligen-fraguas-1993,
kuznetsov-shchepetov-1997,fischer-2012}. A theoretical study of
expansions of magnetic fields in spherical, cylindrical, and toroidal
harmonics was given in Ref.~\cite{gambini-2018}. In that study,
toroidal-harmonic expansions for ideal toroidal magnets were discussed,
and numerical fitting of order ten toroidal harmonics was carried out 
for the situation in which fields could be sampled uniformly in the
angles of the toroidal coordinate system. To our knowledge, our work is
the first to address the use of thousands of toroidal harmonics in
fitting nearly uniform magnetic fields to high precision. The methods
that we employ do not require uniform sampling of the fields in the
angles of the toroidal coordinate
system.

The expansion of a vacuum magnetic field in toroidal harmonics is
discussed in detail in Ref.~\cite{milligen-fraguas-1993}. Here, we
present the essential features of that expansion.

\begin{figure}[htbp]
\centering
\includegraphics[width=0.45\columnwidth]{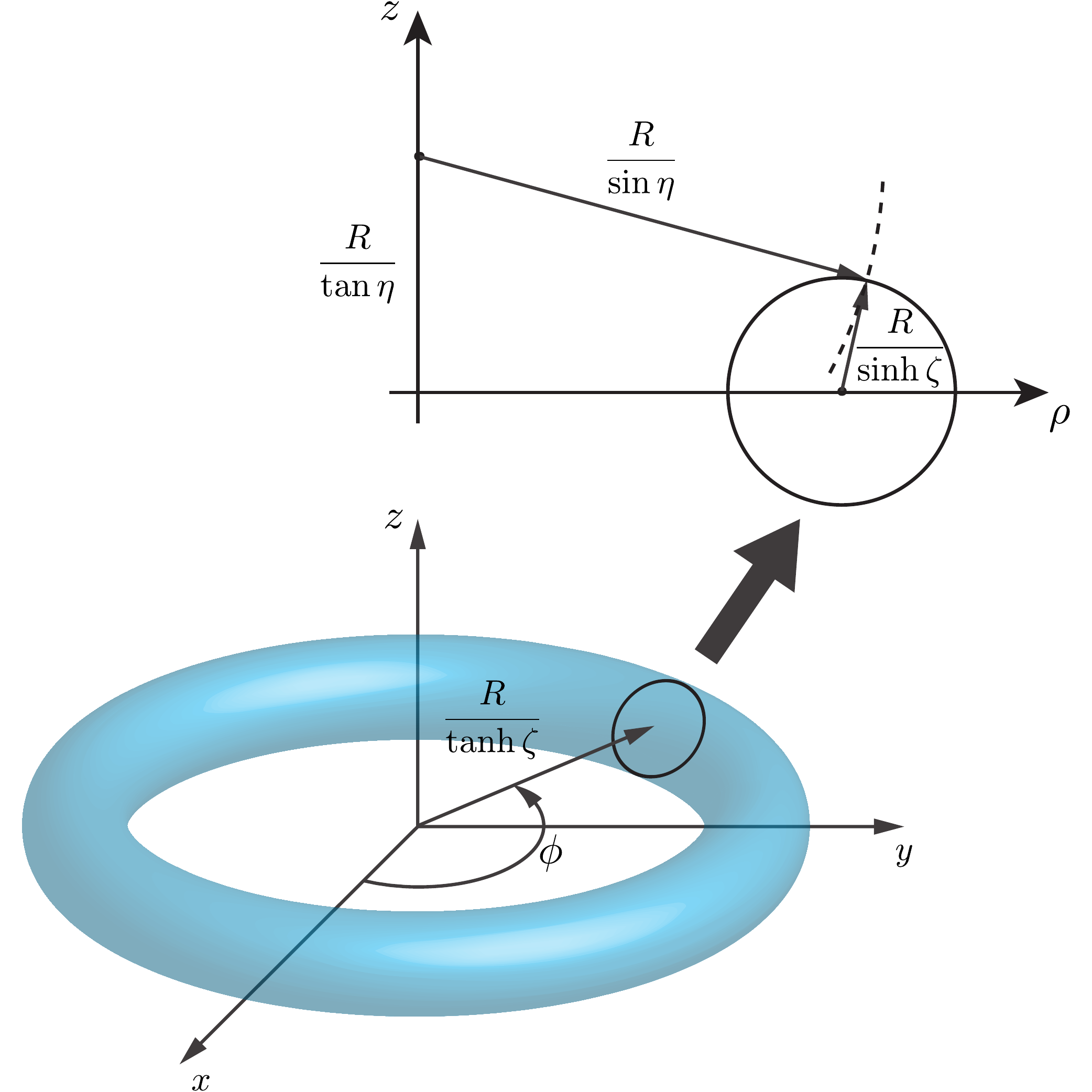}
\caption{Illustration of toroidal coordinates. The surface of constant
$\zeta$ is a† torus of major and minor radii of $R/\tanh \zeta$ and
$R/\sinh \zeta$, respectively, whose center is located at $x=y=z=0$. The
surface of constant† $\eta$ is a sphere of radius $R/\sin \eta$, whose
origin is at $x=y=0$ and† $z=R/\tan \eta$.
\label{fig:toroidal-coord}}
\end{figure}

Toroidal coordinates $(\zeta,\eta,\phi)$ are given in terms of
Cartesian coordinates $(x,y,z)$ as follows:
\begin{subequations}
\begin{eqnarray}
x &=& R \frac{\sinh \zeta}{\cosh \zeta - \cos \eta} \cos \phi, 
\\
y &=& R \frac{\sinh \zeta}{\cosh \zeta - \cos \eta} \sin \phi, 
\\
z &=& R \frac{\sin \eta}{\cosh \zeta - \cos \eta}. 
\end{eqnarray}
\end{subequations}
Here, $0 \leq \zeta < \infty$, $0 \leq \eta < 2 \pi$, and $0 \leq
\phi < 2 \pi$. $\phi$ is the usual azimuthal angle about the $z$
axis, with $\phi=0$ corresponding to the $x$ axis. 
$R$ is a constant that has a dimension of length. The
surface of constant $\zeta$ is a torus of major and minor radii of
$R/\tanh \zeta$ and $R/\sinh \zeta$, respectively. ($\zeta \to \infty$
corresponds to a circle of radius $R$ in the $x-y$ plane.
$\zeta\to 0$ corresponds to the $z$ axis for $\eta$ nonzero and to one
or more of $x$, $y$, and $z$ becoming infinite when $\eta$ goes to zero,
depending on the relative speeds with which $\zeta$ and $\eta$ go to
zero.) The poloidal angle $\eta$ specifies, for each circular $\phi$
slice of the torus, the position on the circle. The surface of constant¬†
$\eta$ is a sphere of radius $R/\sin \eta$, whose origin is at $x=y=0$
and $z=R/\tan \eta.$ The toroidal coordinates are illustrated in
figure~\ref{fig:toroidal-coord}.

Although there is some freedom in choosing the value of $R$, it should
not be too different from the radius of the storage ring, so that a
torus defined by a constant $\zeta$ roughly coincides with the interior
of the storage ring. For numerical calculations, $R$ should be adjusted
so that the points of interest do not lie exactly on the circle of
radius $R$ in the $x-y$ plane; otherwise, one encounters $\zeta =
\infty$ in coordinate transformations. In practice, we take $R$ to be 
0.99993 times the $g-2$ ring radius, which corresponds to a shift of 
0.5~mm from the ring center. This leads to a maximum value of 
$\zeta=10.26$ for the central NMR probe in the $g-2$ trolley. That value 
is easily accommodated by the available numerical dynamic range in our 
computation. We have tried varying $R$ in the range 0.997118--1.00274 
times the $g-2$ ring radius, and find that that variation affects our results 
at the level of a few ppb or less, which is negligible in comparison with other 
uncertainties.

The general solution to Laplace's equation, which is found by using
the technique of separation of variables in toroidal coordinates, is
a linear combination of the following toroidal harmonics, which are defined for
nonnegative integers $m$ and $n$ by
\begin{subequations}
\begin{eqnarray}
&& \sqrt{\cosh \zeta - \cos \eta} \, 
P_{m-1/2}^n (\cosh \zeta) \cos (m \eta) \cos (n \phi), \\
&& \sqrt{\cosh \zeta - \cos \eta} \, 
P_{m-1/2}^n (\cosh \zeta) \cos (m \eta) \sin (n \phi), \\
&& \sqrt{\cosh \zeta - \cos \eta} \, 
P_{m-1/2}^n (\cosh \zeta) \sin (m \eta) \cos (n \phi), \\
&& \sqrt{\cosh \zeta - \cos \eta} \, 
P_{m-1/2}^n (\cosh \zeta) \sin (m \eta) \sin (n \phi), \\
&& \sqrt{\cosh \zeta - \cos \eta} \, 
Q_{m-1/2}^n (\cosh \zeta) \cos (m \eta) \cos (n \phi), \\
&& \sqrt{\cosh \zeta - \cos \eta} \, 
Q_{m-1/2}^n (\cosh \zeta) \cos (m \eta) \sin (n \phi), \\
&& \sqrt{\cosh \zeta - \cos \eta} \, 
Q_{m-1/2}^n (\cosh \zeta) \sin (m \eta) \cos (n \phi), \\
&& \sqrt{\cosh \zeta - \cos \eta} \, 
Q_{m-1/2}^n (\cosh \zeta) \sin (m \eta) \sin (n \phi).
\end{eqnarray}
\end{subequations}
Here, the toroidal functions 
$P_{m-1/2}^n (\cosh \zeta)$ and $Q_{m-1/2}^n (\cosh \zeta)$ are
half-integer Legendre functions of the first and second kind, respectively. 
$P_{m-1/2}^n (\cosh \zeta)$ and $Q_{m-1/2}^n (\cosh \zeta)$ can be
expressed in terms of the hypergeometric function as
\cite{milligen-fraguas-1993,abramovitz-stegun-1972}
\begin{subequations}
\label{toroidal-hypergeometric}
\begin{eqnarray}
P_{m-1/2}^{n} (\cosh \zeta) &=&
\frac{\Gamma(m+n+\tfrac{1}{2})}{2^n n! \Gamma(m-n+\tfrac{1}{2})} 
\tanh ^n \zeta \cosh ^{m-1/2} \zeta 
\nonumber \\ && \times 
\,{}_2F_1 [ \tfrac{1}{2} (n-m+\tfrac{1}{2}), 
\tfrac{1}{2} (n-m+\tfrac{3}{2}); 
n+1; \tanh^2 \zeta ], 
\\
Q_{m-1/2}^{n} (\cosh \zeta) &=&
\sqrt{\frac{\pi}{2}} 
\frac{\Gamma(m+n+\tfrac{1}{2})}{2^m \Gamma(m+1)} 
\frac{\tanh^n \zeta}{\cosh^{m+1/2} \zeta} 
\nonumber \\ && \times 
\,{}_2F_1 [ \tfrac{1}{2} (m+n+\tfrac{1}{2}),\tfrac{1}{2} (m+n+\tfrac{3}{2});
m+1;{\rm sech}^2 \zeta]. 
\end{eqnarray}
\end{subequations}
The $P_{m-1/2}^{n} (\cosh \zeta)$ correspond to the solutions that are
singular at $\zeta \to \infty$, and the $Q_{m-1/2}^{n} (\cosh \zeta)$
correspond to the solutions that are regular at $\zeta \to \infty$.
Therefore, harmonics containing $P_{m-1/2}^n (\cosh \zeta)$ correspond to
contributions from sources in the interior of the ring, and harmonics
containing $Q_{m-1/2}^n (\cosh \zeta)$ correspond to contributions from
sources outside the ring. We use the forms in
eq.~(\ref{toroidal-hypergeometric}) for our numerical
computations.\footnote{We have tested the Mathematica built-in
hypergeometric functions numerically against the algorithm in the
appendix of Ref.~\cite{milligen-fraguas-1993}, for which truncation
errors are given, and find good agreement over the range of indices $m$ 
and $n$ that we use.}

In practical calculations, it is necessary to work with a finite number
of harmonics. This may result in significant inaccuracies if the
potential $V$ is not single valued\footnote{For example, the Fourier
expansion for the function $f(\theta) = \theta$ for $0 \leq \theta \leq
2 \pi$ does not converge uniformly.}. The potential $V$ may not be
single valued when the region of space that satisfies
eq.~\eqref{eq:laplace} is multiply connected. For example, the condition
$V(\zeta, \eta + 2 \pi, \phi) = V(\zeta, \eta, \phi)$ can be spoiled if
there is a current that flows inside the ring because the line integral
$\oint_C d\bm{x} \cdot \bm{B}$ along a closed curve $C$ does not vanish
if $C$ encloses a nonzero net current. For the muon storage ring, the
condition $V(\zeta, \eta + 2 \pi, \phi) = V(\zeta, \eta, \phi)$ holds
because, in the absence of the muon beam itself, the interior of the ring
is a vacuum. The condition $V(\zeta, \eta, \phi + 2 \pi) = V(\zeta,
\eta, \phi)$ is ensured by the fact that $\oint_C d\bm{x} \cdot \bm{B} =
0$, where the path $C$ is on the interior of the ring and encloses the
$z$ axis. The vanishing of this line integral follows from the fact that
no net current flows through the plane of the ring at radii that
lie inside the outermost part of the vacuum chamber.

We set the cutoffs for $m$ and $n$ to be $M$ and $N$, respectively. In
order to eliminate solutions that correspond to currents that pass
through the plane of the storage ring at radii that lie inside the
outermost part of the vacuum chamber, we exclude the harmonics
$P_{m-1/2}^n (\cosh \zeta)$ from the solution. Then, we obtain
\begin{eqnarray}
\label{eq:laplace_master}%
V(\zeta, \eta, \phi) &=& 
\sqrt{\cosh \zeta - \cos \eta} 
\sum_{m=0}^M \sum_{n=0}^N
\frac{Q_{m-1/2}^n (\cosh \zeta) }{
Q_{m-1/2}^n (\cosh \zeta_0) }
\nonumber \\ && \hspace{2ex} \times 
\left[ 
C_{\rm CC}^{mn} \cos (m \eta) \cos (n \phi) 
+C_{\rm CS}^{mn} \cos (m \eta) \sin (n \phi) 
\right.
\nonumber \\ && \hspace{5ex} 
\left.
+C_{\rm SC}^{mn} \sin (m \eta) \cos (n \phi) 
+C_{\rm SS}^{mn} \sin (m \eta) \sin (n \phi) 
\right], 
\end{eqnarray}
where, for numerical stability, we divide each toroidal harmonic by
$Q_{m-1/2}^n (\cosh \zeta_0)$, so that $Q_{m-1/2}^n (\cosh \zeta) /
Q_{m-1/2}^n (\cosh \zeta_0)$ is always order one when $\zeta \approx
\zeta_0$. This approach allows us to avoid the computation of gamma
functions with large arguments in eq.~(\ref{toroidal-hypergeometric}).
There is some arbitrariness in the choice of $\zeta_0$.  An appropriate
choice for $\zeta_0$ is one such that $R/\tanh \zeta_0$ and $R/\sinh
\zeta_0$ are approximately equal to the major and minor radii,
respectively, of the storage ring.

Once we have fixed the 
$(2 M+1) \times (2 N+1)$ constants\footnote{Since $\sin 0 = 0$, 
$C_{\rm CS}^{m0}$, $C_{\rm SC}^{0n}$,
$C_{\rm SS}^{0n}$ and 
$C_{\rm SS}^{m0}$ 
do not appear in eq.~\eqref{eq:laplace_master}
}
$C_{\rm CC}^{mn}$, $C_{\rm CS}^{mn}$, $C_{\rm SC}^{mn}$, 
and $C_{\rm SS}^{mn}$, we can give an 
approximate reconstruction of the magnetic field in the storage ring.
The magnetic field components are given by the gradient of
eq.~\eqref{eq:laplace_master}. The magnetic field can be written
in terms of cylindrical basis vectors as 
\begin{equation}
\bm{B} = \langle B \rangle \hat{\bm{z}} + \bm{\nabla} V = 
B_\rho \hat{\bm{\rho}} + B_z \hat{\bm{z}} + B_\phi \hat{\bm{\phi}}, 
\end{equation}
where \cite{milligen-fraguas-1993} 
\begin{subequations}
\label{eq:mfieldcomps}%
\begin{eqnarray}
B_\rho &=& 
\frac{\sinh \zeta}{R} 
\left( 
\frac{1-\cosh \zeta \cos \eta}{\sinh \zeta} \frac{\partial V}{\partial \zeta}
- \sin \eta \frac{\partial V}{\partial \eta} \right), 
\\
B_z &=& 
\langle B \rangle + 
\frac{\sinh \zeta}{R} 
\left( 
- \sin \eta \frac{\partial V}{\partial \zeta} 
- \frac{1-\cosh \zeta \cos \eta}{\sinh \zeta} \frac{\partial V}{\partial
  \eta}
\right),\label{eq:Bz}
\\
B_\phi &=& \frac{\cosh \zeta - \cos \eta}{R \sinh \zeta} 
\frac{\partial V}{\partial \phi}. 
\end{eqnarray}
\end{subequations}
Hereafter, we refer to $B_\rho$, $B_z$, and $B_\phi$ as 
radial, vertical, and longitudinal fields, respectively. 

One might ask whether other coordinate systems in which Laplace's 
equation is separable, such as spherical coordinates or cylindrical 
coordinates, could be used to analyze the magnetic field of the $g-2$ 
ring. In principle, such coordinate systems are also valid for this 
purpose. However, the resulting mis-match between the coordinate 
system and the ring geometry would make the expansions in the 
corresponding harmonics converge slowly.

One might also ask whether one should sample the magnetic field 
uniformly over the toroidal coordinate system or uniformly over the 
plane of each slice in azimuthal angle, as is done in the $g-2$ 
experiment. While uniform sampling over the toroidal coordinate system 
might lead to some simple relationships between the sample values and the 
constants $C_{\rm CC}^{mn}$, $C_{\rm CS}^{mn}$, $C_{\rm SC}^{mn}$, 
$C_{\rm SS}^{mn}$, it would not reduce the number of harmonics that 
are required to achieve a given precision. More importantly, for the 
$g-2$ experiment, the key issue is the uniformity of the field over the 
entire fiducial volume, and this is best assessed with a uniform 
sampling of the field over the plane of each slice in azimuthal angle.

\section{Fitting the toroidal-harmonic series to data}
\label{sec:fit}

Now we discuss the strategy to constrain the magnetic field by
comparing the expressions in eqs.~(\ref{eq:laplace_master}) and
(\ref{eq:mfieldcomps}) with the measured magnetic field.

\subsection{Linearization of the problem}

We determine the constants $C_{\rm CC}^{mn}$, $C_{\rm CS}^{mn}$, $C_{\rm
SC}^{mn}$, and $C_{\rm SS}^{mn}$ by fitting the expressions in
eqs.~(\ref{eq:laplace_master}) and (\ref{eq:mfieldcomps}) to the
measured magnetic field strengths $B_{\rm measured}$. The most
straightforward way to do this is to find a set of constants that
minimizes the sum of squares of residuals $S$, which is defined by
\begin{equation}
\label{eq:residuals}%
S = \sum_{p=1}^{N_{\rm data}} 
\left[ B_{\rm measured} (\bm{x}_p) 
- |\langle B \rangle \hat{\bm{z}} + \bm{\nabla} V (\bm{x}_p)| 
\right]^2,
\end{equation}
where $\bm{x}_p$ is the position at which the measurement of the
$p$'th data point is made, $N_{\rm data}$ is the total number of
measurements, and $V$ is given by eq.~\eqref{eq:laplace_master}.

Although minimization of $S$ is a general method with which to fit the
data, it is not feasible to carry out in practice, especially when $M$,
$N$, and $N_{\rm data}$ are large. Each component of $\bm{B} = B_\rho
\hat{\bm{\rho}} + B_z \hat{\bm{z}} + B_\phi \hat{\bm{\phi}}$ at a given
position in space is a linear polynomial that depends on $N_{\rm
constants} \equiv (2 M+1) \times (2 N+1)$ constants. Then, the field
strength $|\bm{B}| = \sqrt{B_\rho^2 + B_z^2 + B_\phi^2}$ that appears in
eq.~\eqref{eq:residuals} is a square root of a quadratic polynomial
that depends on $N_{\rm constants}$ constants, which can contain
up to $\tfrac{1}{2} (N_{\rm constants}+1) (N_{\rm constants}+2)$ terms.
Since the sum of the square roots of different polynomials does
not, in general, simplify, the number of terms in $S$ is about
$\tfrac{1}{2} N_{\rm constants}^2 N_{\rm data}$. In the case of our fits
to the storage-ring field data, $N_{\rm constants}$ is typically of order
$10^3$ and $N_{\rm data}$ is typically of order $10^6$. Hence, the
computation of $S$ becomes very compute-time intensive.

In order to simplify the fitting problem, we take the following approach. 
We expand the magnitude of the field strength 
\begin{equation}
|\bm{B}| = \sqrt{\langle B \rangle^2 + 2 \langle B \rangle \hat{\bm{z}} \cdot
\bm{B}_{\rm f} + \bm{B}_{\rm f}^2}. 
\end{equation}
in powers of $\bm{B}_{\rm f}$:
\begin{equation}
|\bm{B}| = \sqrt{
(\langle B \rangle + \hat{\bm{z}} \cdot
\bm{B}_{\rm f})^2  +
(\hat{\bm{\rho}} \cdot \bm{B}_{\rm f})^2 + 
(\hat{\bm{\phi}} \cdot \bm{B}_{\rm f})^2 
} 
= \langle B \rangle + \hat{\bm{z}} \cdot
\bm{B}_{\rm f} 
+ O\left( \bm{B}_{\rm f}^2/\langle B \rangle \right). 
\end{equation}
Since the transverse components of the magnetic field in the $g-2$
storage ring are typically about 100 ppm, the corrections beyond
linear order are of order 0.01 ppm, which is of the same order as the
nominal error in the NMR-probe measurements. Therefore, we can
reasonably ignore the corrections beyond the linear order. In the linear
approximation, we have $|\bm{B}| = B_z$. Therefore, in the linear
approximation, the sum of squares of residuals becomes
\begin{equation}
\label{eq:residuals_linear}%
S_L = \sum_{p=1}^{N_{\rm data}} 
\left[ B_{\rm measured} (\bm{x}_p) - B_z(\bm{x}_p)
\right]^2.
\end{equation}
$B_z$ is a linear polynomial containing $N_{\rm constants}$
constants. Then, each term in the sum in
eq.~\eqref{eq:residuals_linear} is a quadratic polynomial containing
$N_{\rm constants}$ constants, which can have up to $\tfrac{1}{2}
(N_{\rm constants}+1) (N_{\rm constants}+2)$ terms. When we add
polynomials to perform the sum in eq.~\eqref{eq:residuals_linear}, we
can simply add the coefficients numerically. Thus, the number of terms
in $S_L$ is $\tfrac{1}{2} (N_{\rm constants}+1) (N_{\rm constants}+2)$.
This is a factor of $1/N_{\rm data}$ smaller than the number of terms in
$S$. By minimizing $S_L$, we find an approximate solution to the
following set of equations, which, in general, is overconstrained: 
\begin{subequations}%
\begin{equation}
B_z(\bm{x}_p) = 
B_{\rm measured} (\bm{x}_p), 
\label{eq:overconstained}
\end{equation}
where the measurement index $p$ takes on the values $p = 1, 2, \ldots,
N_{\rm data}$. Since $B_z(\bm{x}_p)$ is a linear polynomial that
contains $N_{\rm constants}$ constants, 
eq.~(\ref{eq:overconstained}) can be rewritten as
\begin{equation}
\sum_{i=1}^{N_{\rm constants}} 
\frac{\partial B_z (\bm{x}_p)}{\partial c_i} c_i =
B_{\rm measured} (\bm{x}_p) - \langle B \rangle.
\label{eq:Bz-equal-Bmeasured2}
\end{equation}
\end{subequations}%
where $c_1, c_2, \ldots, c_{N_{\rm constants}}$ denote the constants 
$C_{\rm CC}^{mn}$, $C_{\rm CS}^{mn}$, $C_{\rm SC}^{mn}$, and $C_{\rm SS}^{mn}$.

A further advantage of the linear approximation is that it
reduces the minimization problem to a linear-algebra problem.
Since $S_L$ is a sum of squares of linear polynomials, it
is bounded from below, and, if $S_L$ has one local minimum, 
then the local minimum
coincides with the global minimum. The condition for a local minimum is
\begin{equation}
\label{eq:localmin}%
\frac{\partial S_L}{\partial c_i} = 0, 
\end{equation}
for all $i=1,2,\ldots,N_{\rm constants}$. 
Since $S_L$ is a quadratic polynomial, the left side of eq.~\eqref{eq:localmin} 
is a linear polynomial. Therefore, eq.~\eqref{eq:localmin} can be rewritten as
\begin{equation}
\label{eq:localminmatrix}%
\sum_{j=1}^{N_{\rm constants}} A_{ij} c_j = b_i,
\end{equation}
where 
\begin{subequations}
\label{eq:abmatrices}%
\begin{eqnarray}
A_{ij} &\equiv& \frac{1}{2}\frac{\partial^2 S_L}{\partial c_i \partial c_j}
=\sum_{p=1}^{N_{\rm data}}\frac{\partial B_z(\bm{x}_p)}{\partial c_i}
\frac{\partial B_z(\bm{x}_p)}{\partial c_j},
\\
b_i &\equiv& - \frac{1}{2}\frac{\partial S_L}{\partial c_i} \bigg|_{c_j=0
\textrm{ for all
}j}=\sum_{p=1}^{N_{\rm data}} [B_{\rm measured}(\bm{x}_p) - \langle B \rangle ]
\frac{\partial B_z(\bm{x}_p)}{\partial c_i}.
\end{eqnarray}
\end{subequations}

The square matrix $A$ and the column matrix $b$ are numerical and do not
depend on the $c$'s. $A$ is symmetric and positive semidefinite.
Eq.~\eqref{eq:localminmatrix} has a unique solution if $A$ is positive
definite, {\it i.e.}, if $A$ has no vanishing eigenvalues. However, as
we will explain in the next section, this is not the case in our
application. That is, our linear least-squares problem is 
underconstrained.

There are many well developed methods for solving
eq.~(\ref{eq:localminmatrix}). One such method is singular-value
decomposition (SVD).\footnote{See, for example,
Ref.~\cite{lawson-hanson-1995}.} In this method, the expression $\partial
B_z(\bm{x}_p)/\partial c_i$ on the left side of
eq.~(\ref{eq:Bz-equal-Bmeasured2}) is regarded as a matrix with indices
$p$ and $i$. Then SVD is used to find the pseudoinverse\footnote{Here,
and throughout this paper, we use the term ``pseudoinverse'' to mean the
Moore-Penrose inverse of a matrix
\cite{moore-1920,bjerhammar-1951,penrose-1955,penrose-1956}. See 
Ref.~\cite{lawson-hanson-1995} for a discussion of the pseudoinverse in
the context of SVD.} of that matrix. A solution to the linear least
squares problem is obtained by applying the pseudoinverse matrix to the
right side of eq.~(\ref{eq:Bz-equal-Bmeasured2}). Again, as we will
explain in the next section, the solution of the linear least-squares
problem is not unique.

The SVD method is particularly well suited to non-unique linear
least-squares problems, such as ours, because it allows one to
identify the zero eigenvalues of the matrix $A$ in
eq.~(\ref{eq:localminmatrix}). However, we have found that the
direct application of the SVD method to solve
eq.~(\ref{eq:localminmatrix}) is impractical for several reasons. First,
the appearance of large numbers of different linear combinations of
unconstrained $c_i$'s makes it difficult to identify those linear
combinations and to sequester them. Second, the computation, when it can
be carried out, is very intensive computationally. Finally, the data
display fluctuations between measurements at different azimuthal angles
that are much larger than differences between the measurements of the
various probes at a fixed azimuthal angle. In order to deal with these
issues, we make a further rearrangement of the linear least-squares
problem. In this re-arrangement, which we describe in detail in
section~\ref{sec:further-rearrangement}, we first analyze the azimuthal
dependence of the data, using Fourier decomposition, and then 
analyze the $\rho$-$z$ of the Fourier coefficients in order to determine 
the toroidal-harmonic coefficients.

\subsection{Underconstrained nature of the linear least-squares problem
\label{sec:underconstrained}}

In the linear approximation, the measured magnetic fields constrain
only $B_z$ directly.  $B_\rho$ and $B_\phi$ are constrained
indirectly from knowledge of $B_z$ and the vacuum Maxwell's equations.
From 
$\bm{\nabla} \cdot \bm{B} = 0$ and  
$\bm{\nabla} \times \bm{B} = \bm{0}$, we obtain the following relations
\begin{subequations}%
\label{maxwells-constraints}
\begin{eqnarray}
&&
\frac{\partial B_z}{\partial z} = 
- \frac{\partial B_\rho}{\partial \rho}
- \frac{1}{\rho} B_\rho  
- \frac{1}{\rho} \frac{\partial B_\phi}{\partial \phi}, 
\\
&&
\frac{\partial B_z}{\partial \rho} = \frac{\partial B_\rho}{\partial z},
\label{eq:curl-z-rho}\\
&&
\frac{\partial B_z}{\partial \phi} = \rho \frac{\partial B_\phi}{\partial z}.
\end{eqnarray}
\end{subequations}
Hence, once $B_z$ is determined, the vertical gradients of the
radial and longitudinal fields can be constrained. However,  aside
from the combination $\partial B_\rho/\partial \rho + \rho^{-1} B_\rho
+ \rho^{-1} \partial B_\phi/\partial \phi$, neither the field
gradients in the radial or the longitudinal directions nor the
absolute sizes of $B_\rho$ and $B_\phi$ can be determined from
$B_z$. This implies that a comparison of the expression for $B_z$ with
the measured magnetic field strengths cannot constrain all $(2 M+1)
\times (2 N+1)$ constants $C_{\rm CC}^{mn}$, $C_{\rm CS}^{mn}$, $C_{\rm
SC}^{mn}$, and $C_{\rm SS}^{mn}$. We will deal with the underconstrained
nature of the fits by making use of singular-value decomposition (SVD)
to solve eq.~(\ref{eq:Bz-equal-Bmeasured2}). In this method, the
unconstrained modes correspond to zero eigenvalues of the matrix $A$. We
discard the associated eigenvectors from the solution. In practice, this
is done by setting to zero any inverse eigenvalues in the SVD
pseudoinverse matrix that correspond to eigenvalues that are smaller
than  a ``tolerance'' $\epsilon$. While this results in a well-defined
solution to the fitting problem, it is important to recognize that this
is just one of an infinite number of solutions to this underconstrained
problem. That means that this solution must be interpreted with care,
recognizing that only the gradients with respect to $z$ of the radial
field $B_\rho$ and the longitudinal field $B_\phi$ are fully determined
by the constraints that are given in eqs.~(\ref{maxwells-constraints}).

\subsection{Further rearrangement of the linear least-squares
problem\label{sec:further-rearrangement}}

As we have mentioned, even the linearized fitting problem can be
formidable to solve computationally, since it may involve on the order
of $10^5$ data points and $10^4$  harmonics. We can considerably improve
the computational efficiency by carrying out the analysis in steps. The
first step is to express the $\phi$ dependence of the data for each NMR
probe in terms of a Fourier series. The second step is to use the
Fourier-series coefficients for all of the probes to determine the
coefficients of the toroidal harmonics. The first step in this approach
relies on three facts: (1) the data from each NMR trolley probe are a
function of $\phi$ alone (at fixed $\rho$ and $z$), (2) the
toroidal-harmonic expansion of the potential $V$ depends on $\phi$ only
through sine and cosine factors in each harmonic, and (3) $B_z$ does not
involve azimuthal derivatives of $V$.

\subsubsection{Fourier series for the probe data}

In order to display the sine and cosine factors in the harmonics
explicitly, we write 
\begin{subequations}%
\label{eq:VCVS-defns}
\begin{eqnarray}
V_{\rm C} (\zeta, \eta, n) &=& 
\sqrt{\cosh \zeta - \cos \eta} 
\sum_{m=0}^M 
\frac{Q_{m-1/2}^n (\cosh \zeta) }{
Q_{m-1/2}^n (\cosh \zeta_0) }
\left[ 
C_{\rm CC}^{mn} \cos (m \eta) 
+C_{\rm SC}^{mn} \sin (m \eta) 
\right], \nonumber\\
\\
V_{\rm S} (\zeta, \eta, n) &=& 
\sqrt{\cosh \zeta - \cos \eta} 
\sum_{m=0}^M 
\frac{Q_{m-1/2}^n (\cosh \zeta) }{
Q_{m-1/2}^n (\cosh \zeta_0) }
\left[ 
C_{\rm CS}^{mn} \cos (m \eta) 
+C_{\rm SS}^{mn} \sin (m \eta) 
\right],\nonumber\\ 
\end{eqnarray}
\end{subequations}
so that 
\begin{equation}
V (\zeta, \eta, \phi) = 
\sum_{n=0}^N [
V_{\rm C} (\zeta, \eta, n) \cos(n \phi) + 
V_{\rm S} (\zeta, \eta, n) \sin(n \phi) ].
\label{eq:V-sin-cos} 
\end{equation}
Applying eq.~(\ref{eq:Bz}) to eq.~(\ref{eq:V-sin-cos}), we obtain
\begin{equation}
\label{eq:Fourier-series-Bz}
B_z(\zeta,\eta,\phi)= \langle B \rangle + 
\sum_{n=0}^N[B_C(\zeta,\eta,n)\cos(n\phi)
 +B_S(\zeta,\eta,n)\sin(n\phi)],
\end{equation}
where
\begin{subequations}%
\label{eq:BCBS}
\begin{eqnarray}
B_C(\zeta,\eta,n)&=& \frac{\sinh \zeta}{R} 
\left[ 
- \sin \eta \frac{\partial V_C(\zeta,\eta,n)}{\partial \zeta} 
- \frac{1-\cosh \zeta \cos \eta}{\sinh \zeta} 
\frac{\partial V_C(\zeta,\eta,n)}{\partial
  \eta}
\right],\\
B_S(\zeta,\eta,n)&=& \frac{\sinh \zeta}{R} 
\left[ 
- \sin \eta \frac{\partial V_S(\zeta,\eta,n)}{\partial \zeta} 
- \frac{1-\cosh \zeta \cos \eta}{\sinh \zeta} 
\frac{\partial V_S(\zeta,\eta,n)}{\partial
  \eta}
\right].
\end{eqnarray}
\end{subequations}

Now we define a truncated Fourier series for each NMR probe that has the
same form as the truncated Fourier series in
eq.~(\ref{eq:Fourier-series-Bz}):
\begin{equation}
B_{\rm Fourier}(q,\phi;N)= 
\sum_{n=0}^N [ \beta(q,2 n; N) \cos (n \phi) 
+ \beta(q, 2 n+1; N) \sin (n \phi) ]. 
\end{equation}
Here, the index $q$, which runs from $1$ to $N_{\rm probe}$, labels
the NMR probe, whose $\zeta$ and $\eta$ coordinates are $\zeta_q$
and $\eta_q$, respectively. Note that we retain the index $N$ in the
arguments of the coefficients $\beta$ as a reminder that these coefficients
depend on the order of truncation of the Fourier series.

\subsubsection{Using the probe data to constrain the Fourier coefficients}

A standard method for determining the coefficients of a Fourier series
is to make use of the orthonormality relations for sines and cosines. In
our case, application of these orthonormality relations would lead to
the following equations for the Fourier coefficients $\beta(q,n;N)$:
\begin{subequations}
\begin{eqnarray}
\beta(q,0;N) &=& 
\frac{1}{2 \pi} \int_{-\pi}^{+\pi} d \phi \,
B_{\rm Fourier} (q,\phi;N),
\\ 
\beta(q,2n;N) &=& 
\frac{1}{\pi} \int_{-\pi}^{+\pi} d \phi \,
B_{\rm Fourier} (q,\phi;N) \cos (n \phi),
\\ 
\beta(q,2n+1;N) &=& 
\frac{1}{\pi} \int_{-\pi}^{+\pi} d \phi \,
B_{\rm Fourier} (q,\phi;N) \sin (n \phi),
\end{eqnarray}
\end{subequations}
where $n=1,2,\ldots,N$. Approximate values of the coefficients can be
obtained by replacing $B_{\rm Fourier}(q,\phi;N)$ by the
measured magnetic field strengths and by replacing the integrals with
finite sums that approximate the integrals. This method, which is
analogous to numerical integration methods, has an obvious weakness that
the samplings in $\phi$ are limited to the points at which the
measurements were made. Therefore, accuracies of the Fourier
coefficients that are computed by using this method  deteriorate as
$n$ approaches the sampling rate of the data. A further drawback to
this method is that it couples the issue of the numerical accuracy of the
Fourier coefficients to the issue of the stability of the Fourier
coefficients with increasing $N$, making it difficult to assess the
latter.

The method we employ to determine $B_{\rm Fourier}(q,\phi;N)$ is a
least-squares fit. That is, we find the minimum of 
\begin{subequations}%
\begin{eqnarray}
\label{eq:chi}
\chi_q^2 (N) &=& 
\frac{1}{
N_{\phi}}
\sum_{i} \left[
\frac{B_{\rm measured}(\zeta_q,\eta_q,\phi_i)
-\langle B \rangle- B_{\rm Fourier} (q,\phi_i;N)}
{\langle B \rangle} \right]^2\\
&=& 
\frac{1}{
N_{\phi}}
\sum_{i} \left[
\frac{B_{\rm measured}(\zeta_q,\eta_q,\phi_i)
-\langle B \rangle- [\partial B_{\rm Fourier}(q,\phi_i)/\partial 
\beta(q,n;N)]\beta(q,n;N)}
{\langle B \rangle} \right]^2, \nonumber\\ 
\end{eqnarray}
\end{subequations}
where the sum is over all values of $\phi$ at which the measurements
were made, and $N_\phi$ is the number of measurements.\footnote{Note
that the quantity $\chi_q^2(N)$ is {\it not} the usual quantity $\chi^2$
that appears in statistics, as it is normalized to $\langle B\rangle$,
rather than to the experimental error.} In practice we solve this
linear least-squares problem by making use of the SVD method. That is,
we find the pseudoinverse of the matrix
\begin{equation}
\frac{\partial B_{\rm Fourier}(q,\phi_i;N)}{\partial 
\beta(q,n;N)},
\end{equation}
whose indices are $i$ and $n$, and apply it to the column vector $B_{\rm
measured}(\zeta_q,\eta_{q},\phi_i)-\langle B\rangle e_i$, whose
index is $i$. (Here $e$ is a column vector with all unit
entries.) We denote by $\chi_q^{\rm min}(N)$ the minimum value of
$\chi_q(N)$: 
\begin{equation}
\chi_q^{\rm min}(N)=\min_{\beta(q,n;N)} \chi_q(N).
\label{eq:chimin}
\end{equation}
We also define 
\begin{equation}
\chi^2(N)=\frac{1}{N_{\rm probe}}\sum_{q=1}^{N_{\rm probe}} 
[\chi_q^{\rm min}(N)]^2. \label{eq:chiN}
\end{equation}

It is obvious that the minimum of $\chi_q (N)$ will decrease
with increasing $N$. Of course, this does not mean that we can increase
$N$ indefinitely. If $N$ is too large, the Fourier coefficients become
very sensitive to small fluctuations in data, which may be comparable
to the uncertainties in measurement. Furthermore, the resulting Fourier
series $B_{\rm Fourier}(q,\phi;N)$ can exhibit
unphysically large fluctuations at values of $\phi$ for which there are
no data.

In order to deal with these issues, we use the following method to
determine a value of $N$ such that the Fourier series best
approximates the actual magnetic field. We note that, if the 
Fourier coefficients are known exactly, then 
\begin{equation}
\bar \chi^2 (N, N') = 
\frac{1}{2 \pi N_{\rm probe}} 
\sum_{q=1}^{N_{\rm probe}}
\int_{-\pi}^{+\pi} d \phi
 \left[
\frac{
B_{\rm Fourier} (q,\phi;N)
- B_{\rm Fourier} (q,\phi;N')}
{\langle B \rangle} \right]^2, \label{eq:chi-bar}
\end{equation}
will decrease  with decreasing $N-N'=\Delta N$ for fixed $N$, and also
for increasing $N$ with fixed $\Delta N$, as long as $N$ is large
enough. We will apply these criteria to determine the optimal
choice of $N$ for the Fourier analysis.
When we compute $\bar \chi (N, N')$, we make use of Parseval's 
theorem, which yields 
\begin{equation}
\bar \chi^2 (N, N') = 
\frac{1}{2 N_{\rm probe}}
\sum_{q=1}^{N_{\rm probe}}
\left[
\frac{2 \, [\beta(q,0;N)-\beta(q,0;N')]^2 + 
\sum_{n=1}^{n_{\rm max}}
[\beta(q,n;N) -\beta(q,n;N')]^2
}{ \langle B \rangle^2 }
\right],
\end{equation}
where $n_{\rm max} = \max (2 N+1, 2 N'+1)$.

\subsubsection{Constraining the toroidal-harmonic coefficients}

Having determined the coefficients $\beta(q,n;N)$ of the Fourier series
in $\phi$ for each probe, we can now complete the determination of the
toroidal coefficients $C_{\rm CC}^{mn}$, $C_{\rm CS}^{mn}$, $C_{\rm
SC}^{mn}$, and $C_{\rm SS}^{mn}$. (Recall that we denote these
toroidal-harmonic coefficients by $c_i$.) Then we determine the $c_i$ by
minimizing the following quantities for each $n$:
\begin{subequations}%
\begin{eqnarray}
&&\sum_q \left[\beta(q,2n;N)-B_C(\zeta_q,\eta_q,n)\right]^2\\
&&=\sum_q \left\{\beta(q,2n;N)-
[\partial B_C(\zeta_q,\eta_q,n)/\partial c_i]c_i\right\}^2
\end{eqnarray}
\end{subequations}%
and
\begin{subequations}%
\begin{eqnarray}
&&\sum_q \left[\beta(q,2n+1;N)-B_S(\zeta_q,\eta_q,n)\right]^2\\
&&=\sum_q \left\{\beta(q,2n+1;N)-[\partial B_S(\zeta_q,\eta_q,n)/\partial 
c_i]c_i\right\}^2
\end{eqnarray}                                           
\end{subequations}%
where $B_C(\zeta_q,\eta_q,n)$ and $B_S(\zeta_q,\eta_q,n)$
are given by eq.~(\ref{eq:BCBS}).

Again, we use the SVD method. We find the pseudoinverses of the matrices 
\begin{equation}
\label{eq:toroidal-fit-matrix}
\frac{\partial B_{C,S}(\zeta_q,\eta_q,n)}{\partial c_i},
\end{equation}
where the matrices have indices $i$, which runs from $1$ to $2M+1$, and
$q$, which runs from $1$ to $N_{\rm probe}$. We apply the pseudoinverse
of the ``C'' matrix to the column vector $\beta(q,2n;N)$, which has
index $q$; we apply the pseudoinverse of the ``S'' matrix to the
column vector $\beta(q,2n+1;N)$, which also has index $q$. This
linear least-squares problem is underconstrained, for the reasons that
we discussed in section~\ref{sec:underconstrained}. Therefore, the SVD
procedure is very convenient for finding a particular solution.

We must carry out this procedure for each of the $2N+1$ Fourier
components. However, the matrices in eq.~(\ref{eq:toroidal-fit-matrix})
have dimensionality $(2M+1)\times N_{\rm probe}$, while the matrix in
eq.~(\ref{eq:Bz-equal-Bmeasured2}) that appeared in the original linear
least-squares problem for the toroidal coefficients has dimensionality
$(2M+1)(2N+1)\times N_{\rm data}$. Consequently, the determination of
the $c_i$ in this step requires much less computational time than does
the determination of the $c_i$ directly from the original least-squares
problem.

In the current $g-2$ experiment, $N_{\rm probe}$ is either $17$ or $25$.
This number is not sufficiently large to allow us to use the analogue of
$\bar \chi(N,N')$ in eq.~(\ref{eq:chi-bar}) to determine the optimum
value for the number of poloidal terms $M$ in the toroidal harmonic
series. Instead, we vary $M$ over a range of $5$--$8$ for the case of
$25$ probes and check to see if there are noticeable differences in the
predictions of the toroidal-harmonic series for $\bm{B}$.

\section{Application of the toroidal-harmonic method to data}
\label{sec:results}

In this section, we demonstrate the toroidal-harmonic method that we
have described by applying it to magnetic-field data from the $g-2$
experiment. In these applications, we take $\langle B\rangle$,
expressed in units of the NMR frequency, to be 61.8~MHz. We take the
tolerance in the SVD calculations of the toroidal coefficients to be
$\epsilon=10^{-8}$. We have found that varying $\epsilon$ over several
orders of magnitude makes very little numerical difference in the
gradients of the magnetic field that are well constrained by Maxwell's
equations.

We emphasize that the results that we present are for illustrative
purposes only. These results are based upon data that were taken early
in the commissioning of the muon-storage-ring magnetic field, and
they data do not reflect either the quality of the magnetic field or
the quality of the magnetic-field measurements that will ultimately be
used in the experiment.

\subsection{Fourier decomposition}

First we show results from the Fourier decomposition in $\phi$ of the data
from each probe.

\subsubsection{Run~52 data}

We begin by examining data from trolley Run~52
(Ref.~\cite{grange-et-al}), which are based on measurements from 25
NMR probes (located in the ``shimming trolley'') at 7842 azimuthal
angles, for a total of 196050 data points.

\begin{figure}[htbp]
\centering
\includegraphics[width=0.45\columnwidth]{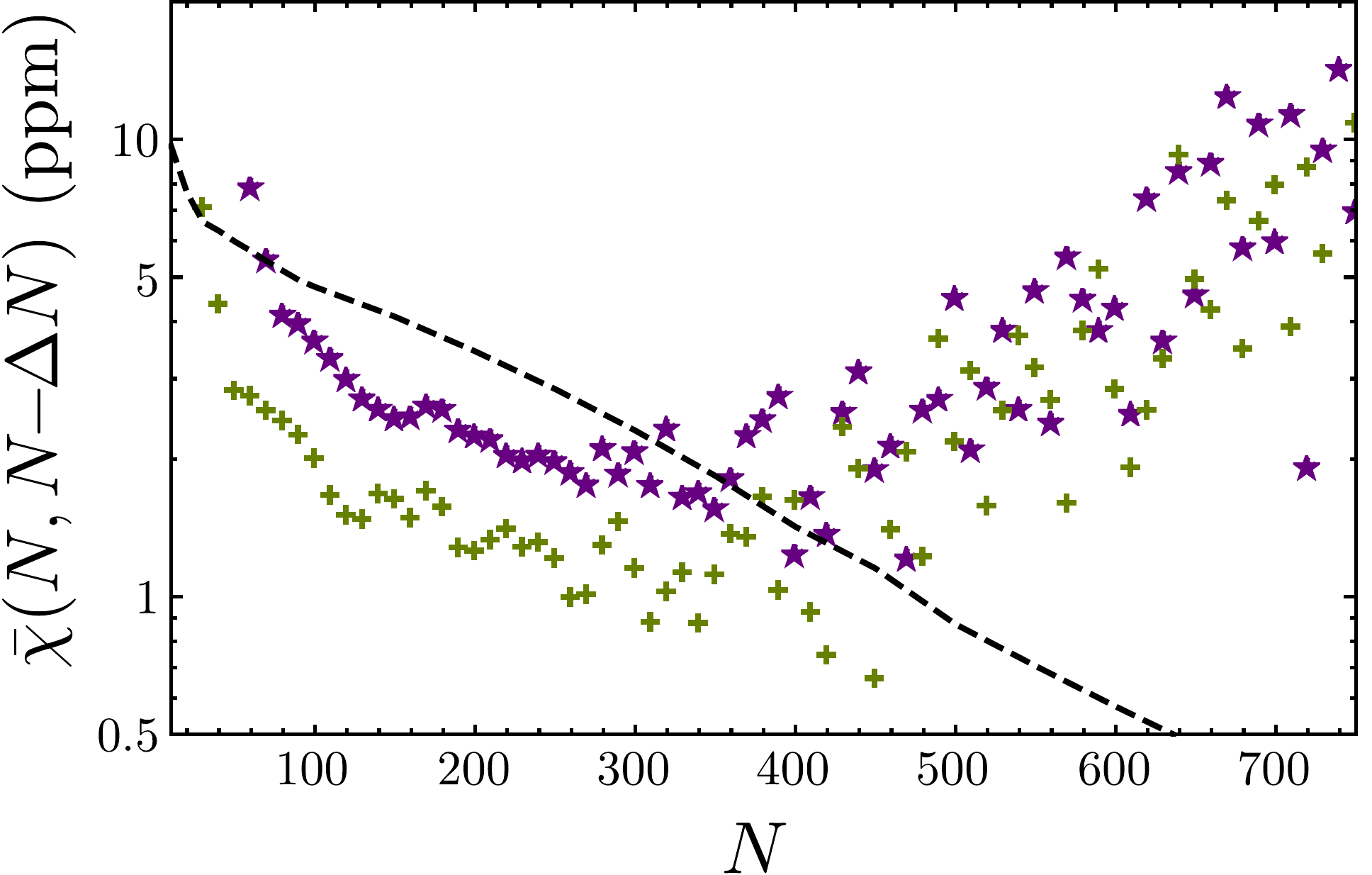}
\includegraphics[width=0.45\columnwidth]{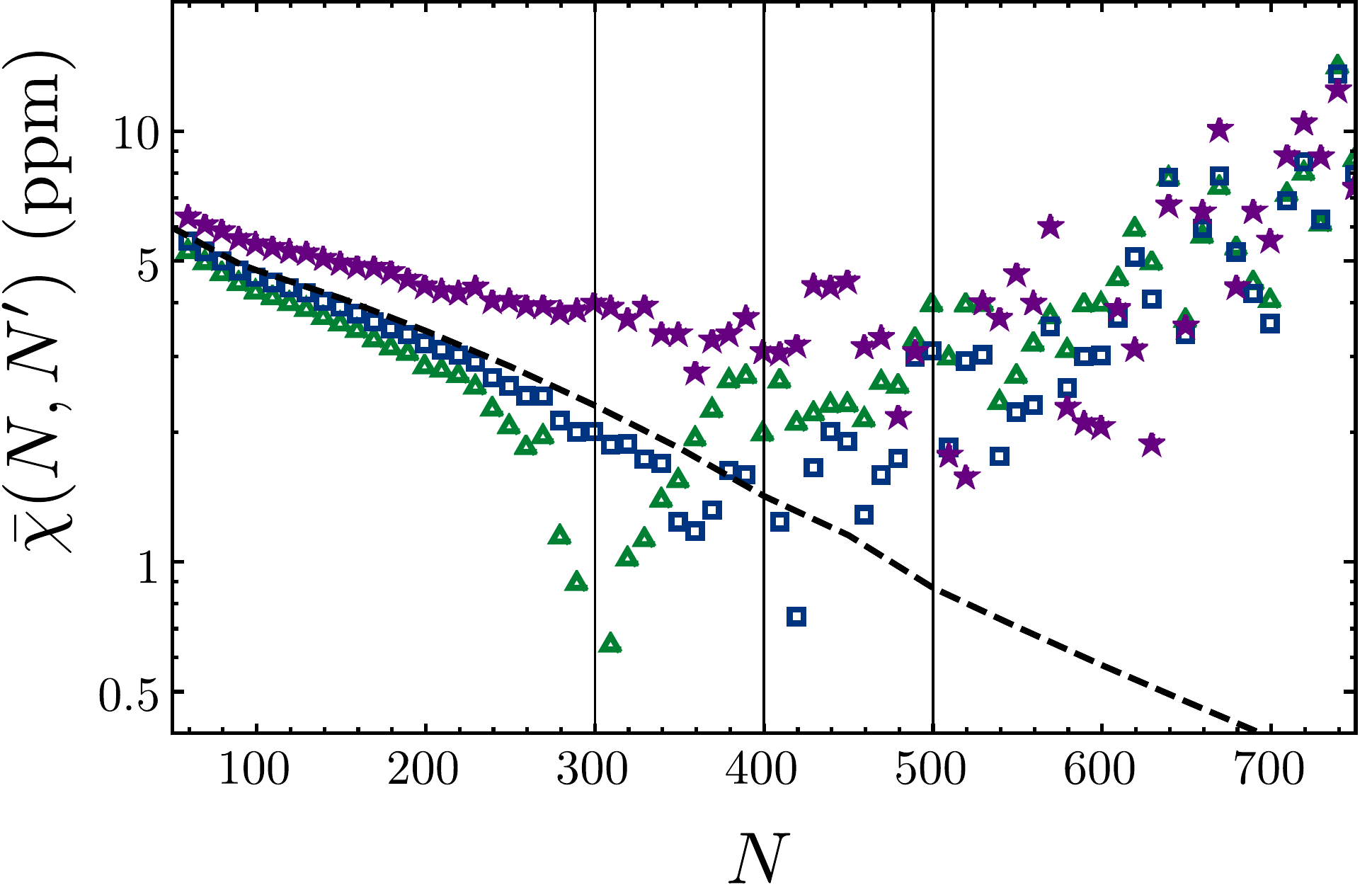}
\caption{Values of $\bar\chi(N,N')$ for the data of Run~52. The left
figure shows $\bar\chi(N,N-\Delta N)$ in units of ppm, as a function
of $N$, with $\Delta N$ fixed to 20 and 50 (shown with crosses and
stars, respectively). The right figure shows  $\bar\chi(N,N')$ in units
of ppm, as a function of $N$,  with $N'$ fixed to 300, 400, and
500 (shown with triangles, squares, and stars, respectively). In both
figures, the dashed line indicates the values of $\chi(N)$ 
[eq.~(\ref{eq:chiN})] in units of ppm.
\label{fig:chi-bar-52}}
\end{figure}

In figure~\ref{fig:chi-bar-52}, we show the quantity $\bar\chi(N,N')$
[eq.~(\ref{eq:chi-bar})].
As can be seen from both of the plots in this figure, the quality of the
Fourier decomposition is optimal around $N=300$. We note that
$\chi(N)$ continues to decrease beyond $N=300$. However, the plots of
$\bar\chi(N,N')$ and $\bar\chi(N,N-\Delta N)$ indicate that the fits
become unstable when $N$ becomes significantly greater then $300$. This
instability manifests itself in large changes in the low-order Fourier
coefficients and in unphysically large fluctuations in the values of the
truncated Fourier series at points that lie between the data points.
Therefore, we choose $N=300$ in our fits, which implies that there are
601 Fourier components in our Fourier analysis of the $\phi$ dependence
for each probe. We have carried out the computation of the Fourier
coefficients using Mathematica code \cite{mathematica-11}. In the
Mathematica code, it is very important to vectorize the matrix
operations, which results in about an order-of-magnitude reduction in
the computational time and almost a factor of 2 reduction in the memory
usage. The computation of the Fourier coefficients takes approximately 4
minutes on a dual 6-core 2.6~GHz Intel Xeon CPU system and consumes
about 1.5~GB of memory in total.

\begin{figure}[htbp]
\centering
\includegraphics[width=0.9\columnwidth]
{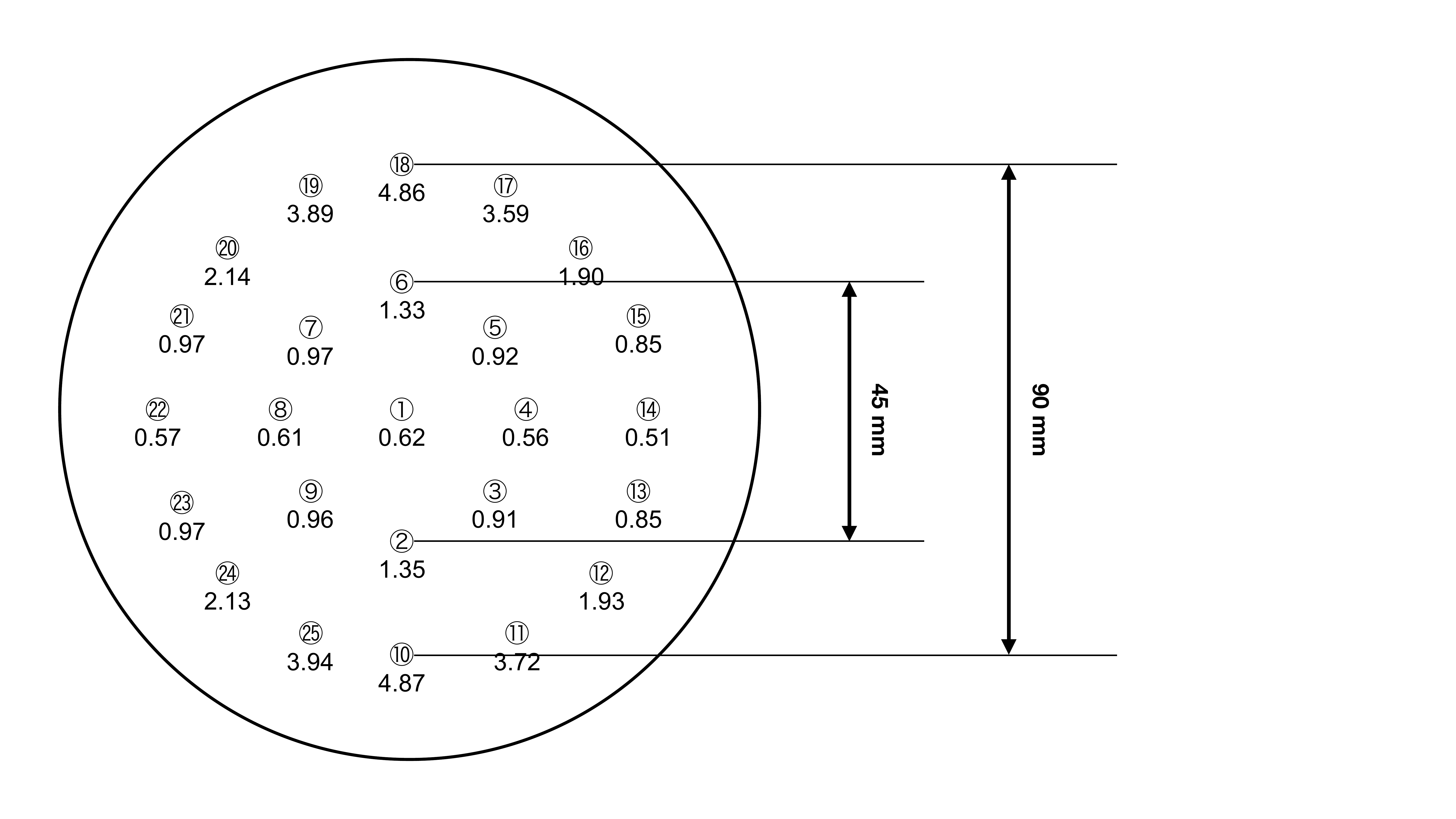}
\caption{Approximate positions in the $\rho$-$z$ plane of the NMR probes
in the 25-probe trolley. The coordinate $\rho$ increases as one 
moves to the right in the figure, and the coordinate $z$ increases as
one moves upward in the figure. The position of probe $1$ is given by
$\rho=7112$~mm and $z=0$~mm. Under each probe position, we list the
value of $\chi_q^{\rm min}(300)$ in ppm that we have obtained for that
probe from the Run~52 data.
\label{fig:probes-chi-52}}
\end{figure}

Already, at the level of the Fourier analysis, the fit provides useful
information about the field measurements. Figure~\ref{fig:probes-chi-52}
shows the approximate positions in the $\rho$-$z$ plane of the 25 NMR
probes. Under each probe position, we list the value of $\chi_q^{\rm
min}(300)$ [eq.~(\ref{eq:chimin})] in ppm that we have obtained for that
probe.
First, we note that the values of $\chi_q^{\rm min}(300)$ that we have
obtained are much larger than the nominal probe measurement uncertainty
of 10~ppb. This suggests that there may be spatial variations in the
magnetic field that are on too small a scale to be represented
accurately by the fit, temporal variations in the magnetic field, and/or
measurement anomalies in the data. We note that the values of
$\chi_q^{\rm min}(300)$ display an interesting spatial distribution:
$\chi_q^{\rm min}(300)$ is larger at the top and bottom of the trolley
than in the middle and is fairly uniform across a given horizontal row.
This may provide a clue as to the origin of the unexpectedly large
values of $\chi_q^{\rm min}(300)$.

\begin{figure}[htbp]
\centering
\includegraphics[width=0.45\columnwidth]{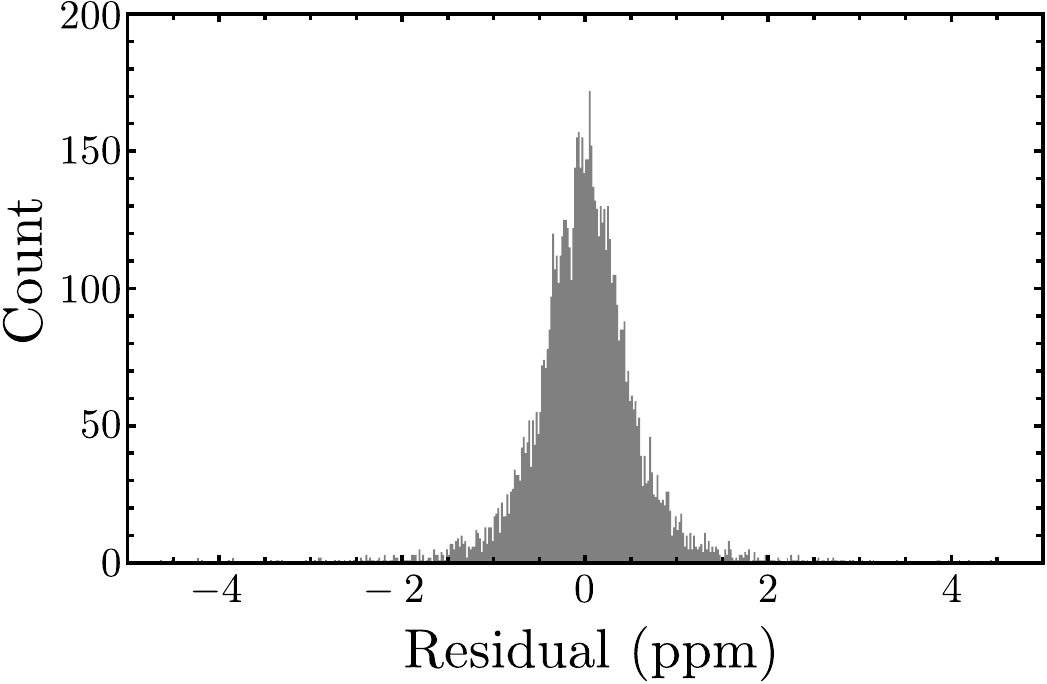}
\includegraphics[width=0.45\columnwidth]{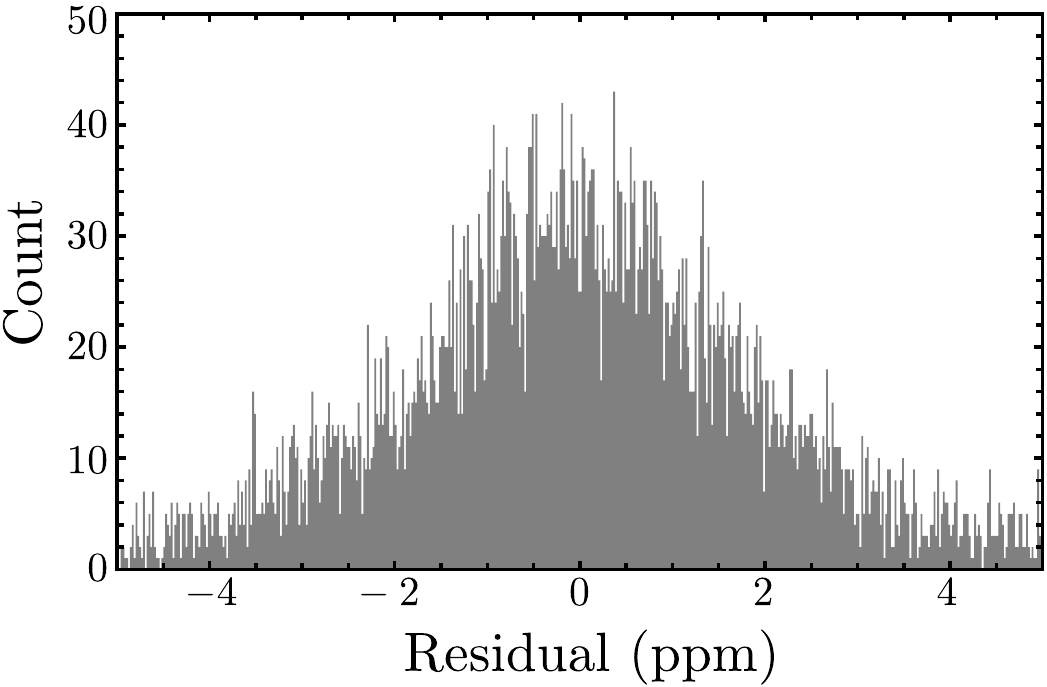}
\caption{Residuals in ppm between the Fourier series fit and the 
data from Run~52 for probe~1 (left figure) and probe~17 (right 
figure). \label{fig:residuals-52}}
\end{figure}

In figure~\ref{fig:residuals-52}, we show the distributions of residuals
between the Fourier series fit and the data from Run~52 for
probes 1 and 17.
As can be seen, the widths of the distributions are rather different, as
would be expected from the different values of $\chi_q^{\rm min}(300)$.

\begin{figure}[htbp]
\centering
\includegraphics[width=0.95\columnwidth]{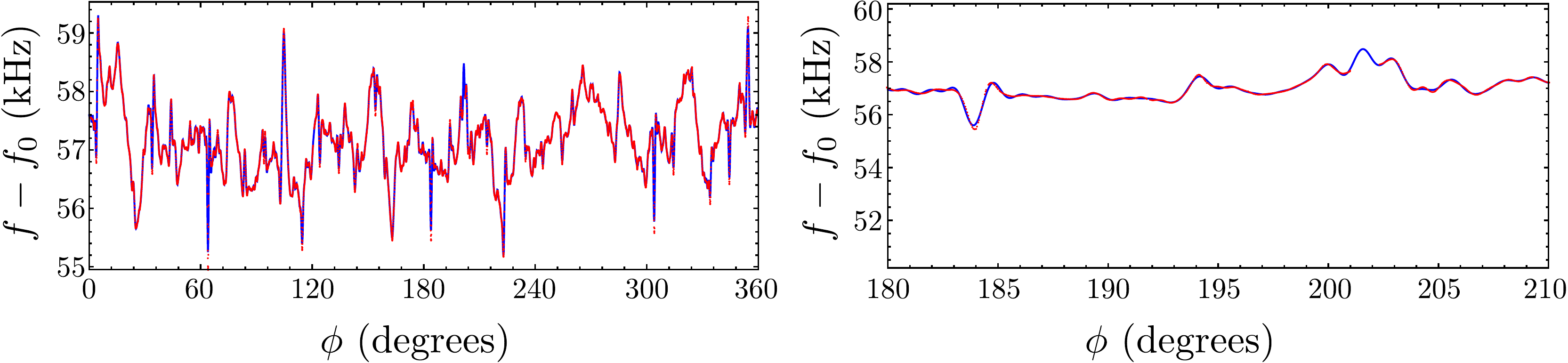}
\includegraphics[width=0.95\columnwidth]{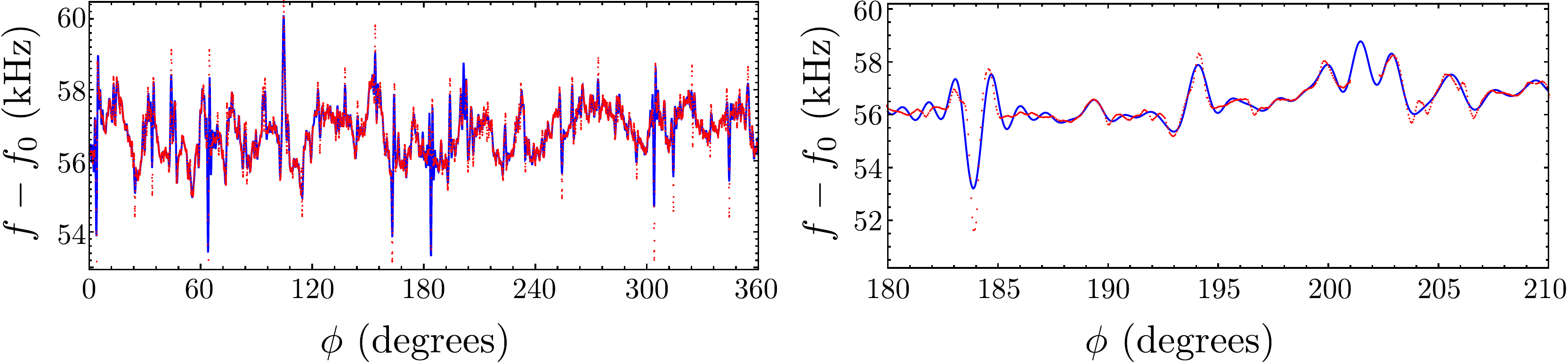}
\caption{Fourier series fit and Run~52 data for probe~1 (upper
figures) and probe~17 (lower figures). The figures on the left show the
full range of $\phi$; the figures on the right are ``zoomed in'' to a
smaller range of $\phi$. The points (red) are the data and the curve 
(blue) is the fit.
The vertical axis is $B_z$ in units of the NMR frequency, 
offset by an amount $f_0=61.74$~MHz. For these data, 1~ppm is 
approximately $0.062$~kHz. 
\label{fig:fitvsdata-52}}
\end{figure}

Figure~\ref{fig:fitvsdata-52} shows the Fourier series fit for probe~1 
and probe~17, along with the data as a function of $\phi$. 
These figures reveal significant deviations of the data from the fit,
especially in the case of probe~17.  One such deviation can be seen
in the ``zoomed-in'' figure for probe~17 near $184^\circ$. There,
the data appear to be smooth in the region of the deviation. However,
the fit does not contain enough harmonics to follow the data accurately.
It is not possible to increase the number of harmonics for this data set
because, as we have mentioned, the fits become unstable as $N$ increases
beyond about $300$. This instability can be traced to nonuniformity
in the sampling of the data over the range of $\phi$. For example, the
gap in the measurements in the range $\phi=201$--$202^\circ$ can cause
instabilities once $N$ is sufficiently large for the harmonics to probe
the gap because the fit function is completely unconstrained in the
range of the gap. We would expect this $1^\circ$ gap to produce
instabilities when $N\sim 360^\circ/1^\circ$, which is in agreement with
the stability analysis that is based on $\bar\chi(N,N-\Delta N)$ and
$\bar\chi(N,N')$.

\subsubsection{Run~3483 data}

\begin{figure}[htbp]
\centering
\includegraphics[width=0.45\columnwidth]{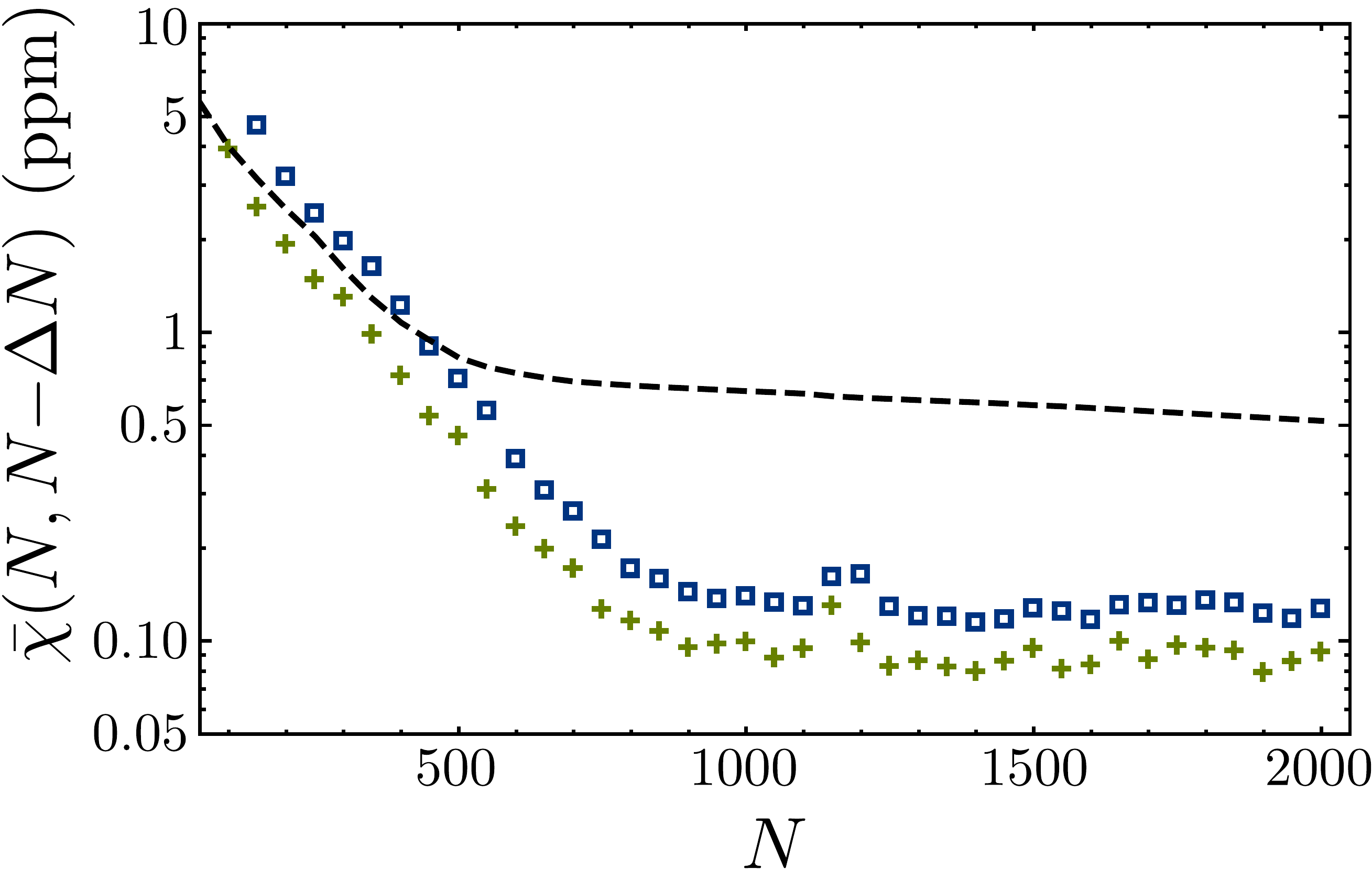}
\includegraphics[width=0.45\columnwidth]{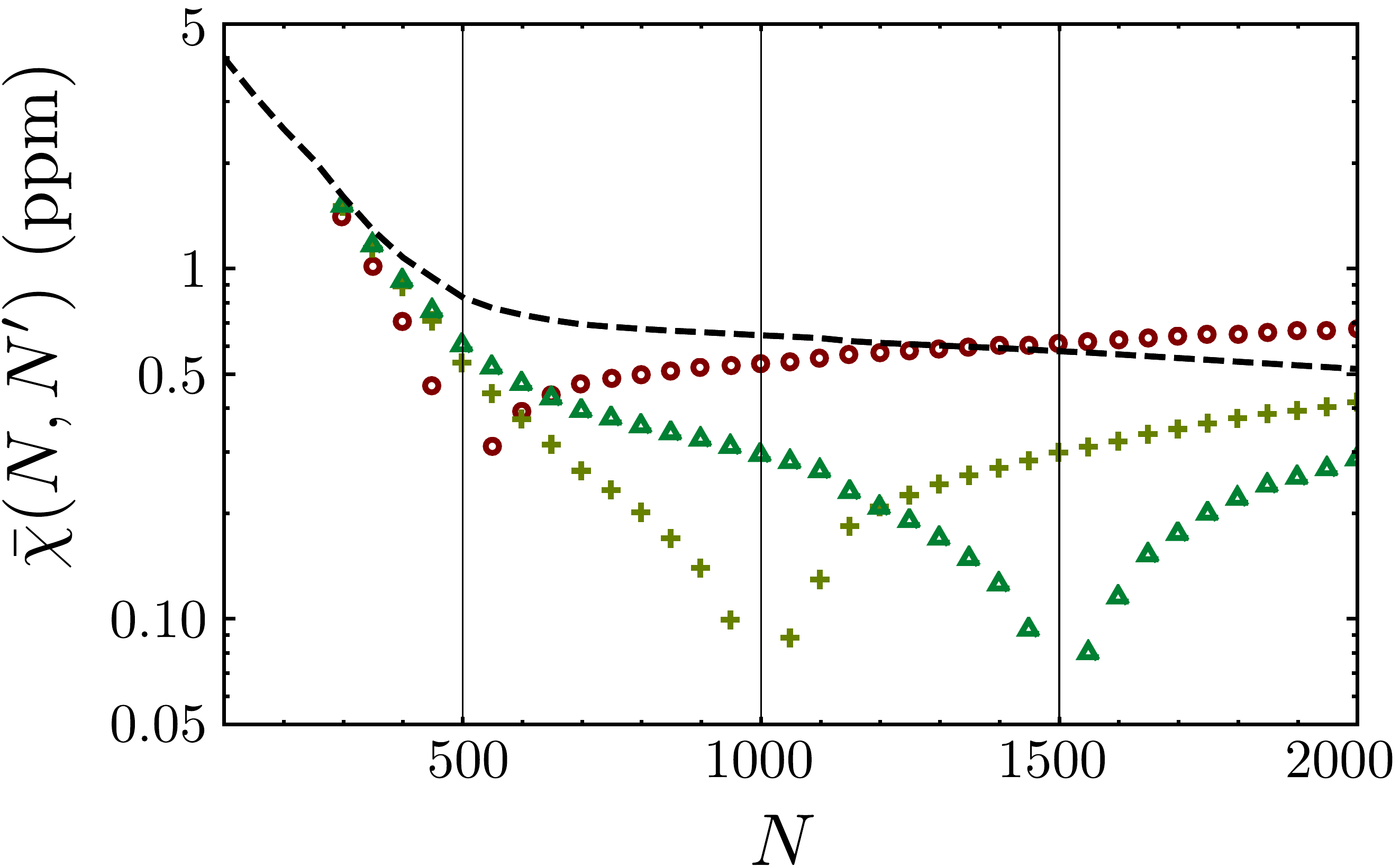}
\caption{Values of $\bar\chi(N,N')$ for the data of Run~3483. The left
figure shows $\bar\chi(N,N-\Delta N)$ in units of ppm, as a function of
$N$, with $\Delta N$ fixed to 50 and 100 (shown with crosses and
squares, respectively). The right figure shows  $\bar\chi(N,N')$ in
units of ppm, as a function of $N$,  with $N'$ fixed to 500, 1000, 1500
(shown with circles, crosses, and triangles, respectively). In both
figures, the dashed line indicates the values of $\chi(N)$
[eq.~(\ref{eq:chiN})] in units of ppm.
\label{fig:chi-bar-3483}}
\end{figure}

For comparison, we also show results from a later trolley run (Run~3483)
in which the trolley contained 17 NMR probes. The data for this run
\cite{grange-et-al} contain measurements at 9023 azimuthal angles,
for a total of 153391 data points. In figure~\ref{fig:chi-bar-3483}, we
show the quantity $\bar\chi(N,N')$ [eq.~(\ref{eq:chi-bar})].
These figures show that $\chi(N)$ changes slope abruptly at $N=500$,
suggesting that most of the features of the data are reproduced by the
fit at that value of $N$. Therefore, we choose $N=500$ for our
fits, which implies that there are $1001$ Fourier components in our
analysis of the $\phi$ dependence for each probe.

\begin{figure}[htbp]
\centering
\includegraphics[width=0.9\columnwidth]
{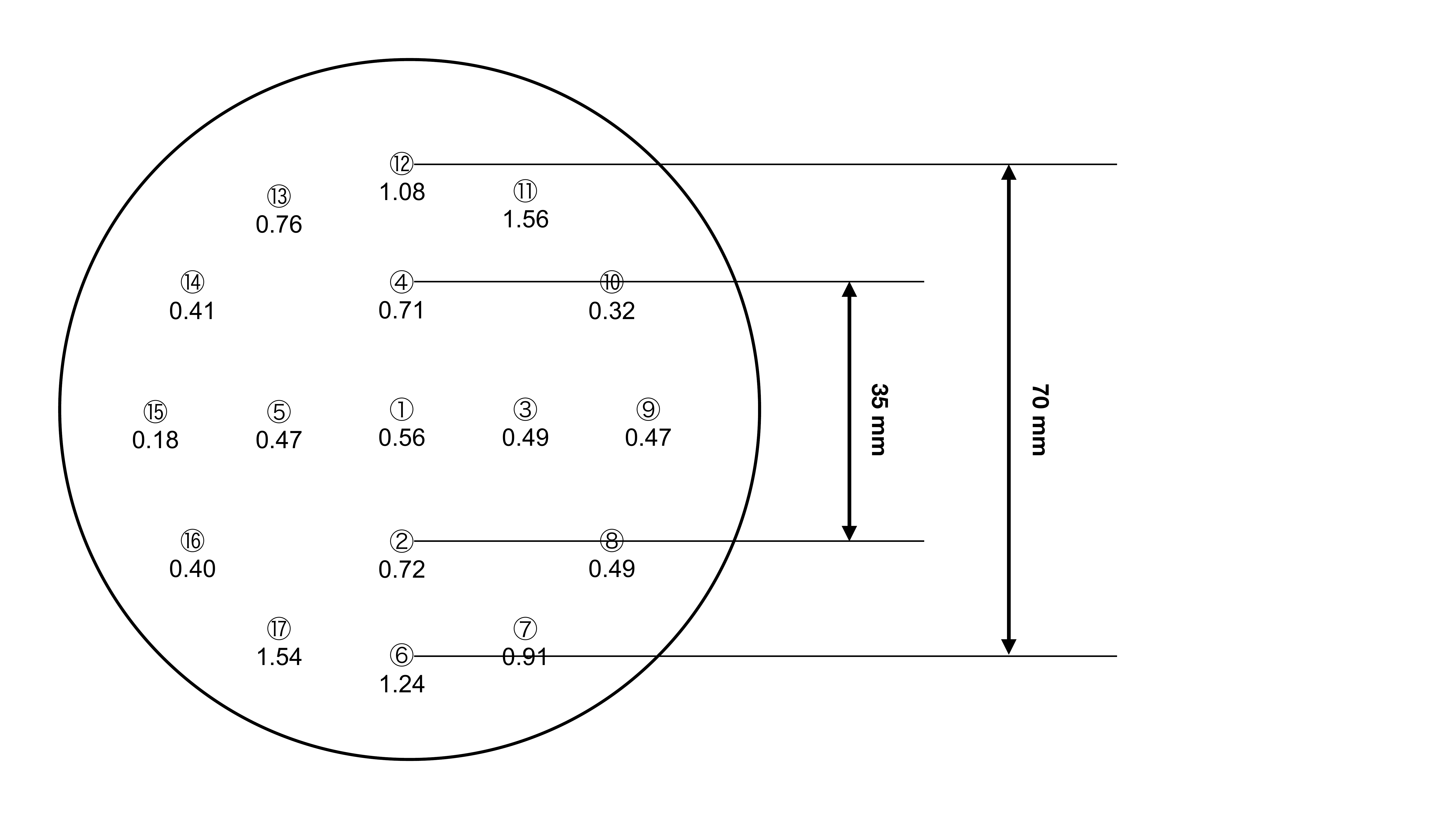}
\caption{Approximate positions in the $\rho$-$z$ plane of the NMR probes
for the 17-probe trolley. The coordinate $\rho$ increases as one 
moves to the right in the figure, and the coordinate $z$ increases as
one moves upward in the figure. The position of probe $1$ is given by
$\rho=7112$~mm and $z=0$~mm. Under each probe position, we list the
value of $\chi_q^{\rm min} (500)$ in ppm that we have obtained
for that probe from the Run~3483 data.
\label{fig:probes-chi-3483}}
\end{figure}

In figure~\ref{fig:probes-chi-3483}, we show the approximate positions in
the $\rho$-$z$ plane of the 17 NMR probes. Under each probe position, we
list the value of $\chi_q^{\rm min}$ [eq.~(\ref{eq:chimin})] in
ppm that we have obtained for that probe.
Again, the values of $\chi_q^{\rm min}(500)$ that we have obtained are
much larger than the nominal probe measurement uncertainty of 10~ppb.
However, they are smaller than the values of $\chi_q$ from Run~52,
perhaps reflecting an improvement in the uniformity of the magnetic
field as a result of the shimming process. The values of $\chi_q^{\rm
min}(500)$ display a spatial distribution that is similar to the spatial
distribution of the values of $\chi_q^{\rm min}(300)$ from Run~52:
The values of $\chi_q^{\rm min}(500)$ are larger at the top and bottom of
the trolley than in the middle. However, the values of $\chi_q^{\rm
min}(500)$ are less uniform across a given horizontal row than are the
values of  $\chi_q^{\rm min}(300)$ from  Run~52.

\begin{figure}[htbp]
\centering
\includegraphics[width=0.45\columnwidth]{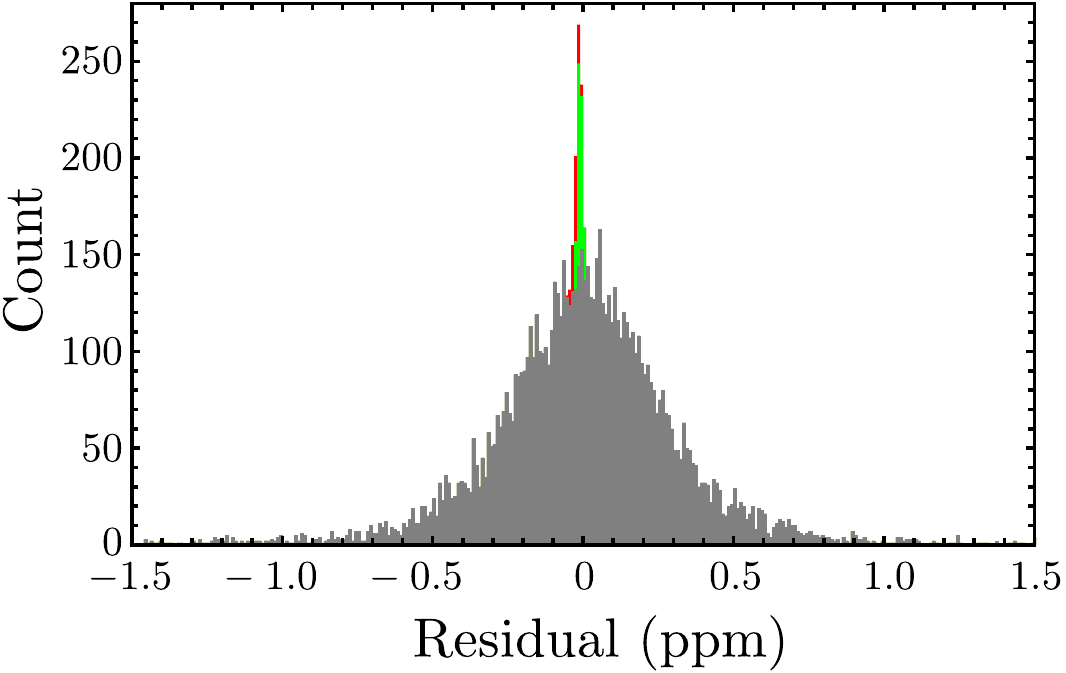}
\includegraphics[width=0.45\columnwidth]{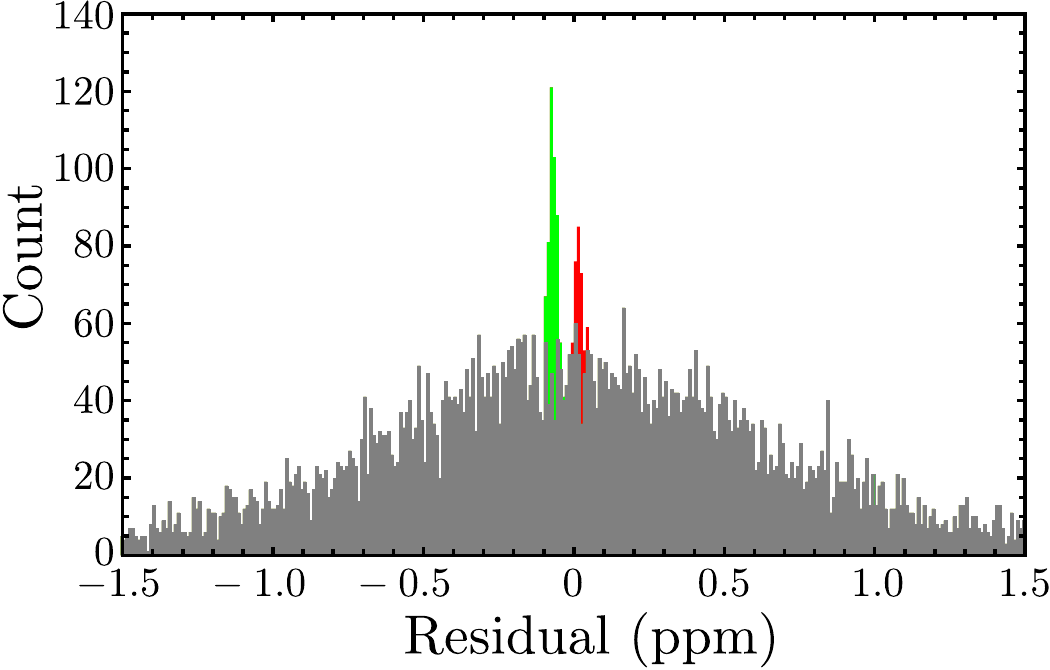}
\caption{Residuals in ppm between the Fourier series fit and the 
data from Run~3483 for probe~1 (left figure) and probe~12 (right 
figure). \label{fig:residuals-3483}}
\end{figure}

In figure~\ref{fig:residuals-3483}, we show the distributions of residuals
between the Fourier series fit and the data from Run~3483 
for probes 1 and 12.
As was the case for the Run~52 data, the widths of the distributions are
rather different. In this case, multiple approximately-Gaussian distributions 
are present. Two narrow distributions (shown in red and light green) arise 
from data in which multiple measurements were taken with the trolley in 
a particular azimuthal location. These distributions reflect the 
reproducibility of the measurements at a given trolley location as a 
function of time. A broader distribution (in gray) arises from 
data taken while the trolley was in motion, and may be indicative of 
actual field variations that are on too small a scale to be represented 
by the fit and/or measurement anomalies.

\begin{figure}[htbp]
\centering
\includegraphics[width=0.95\columnwidth]{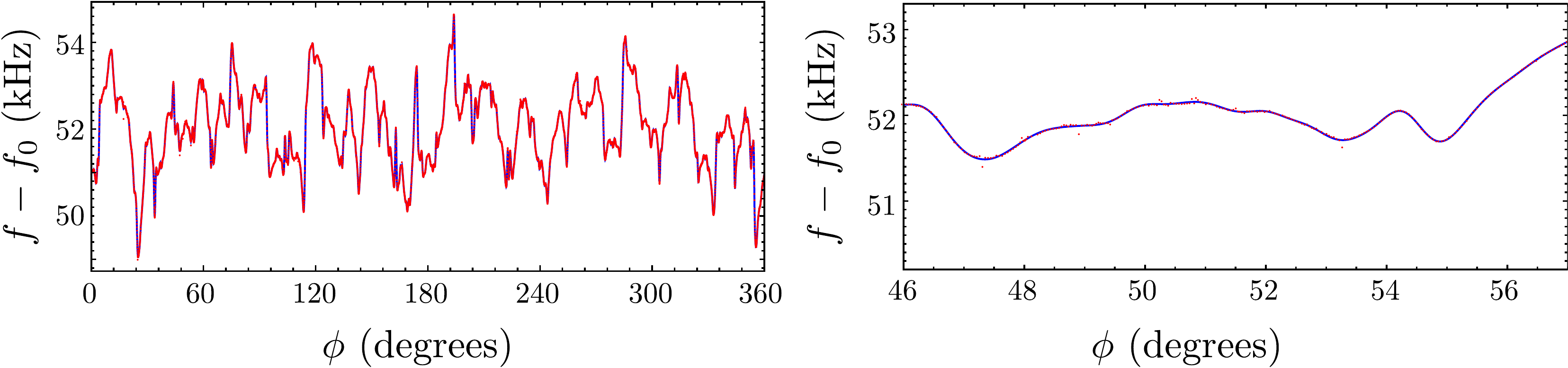}
\includegraphics[width=0.95\columnwidth]{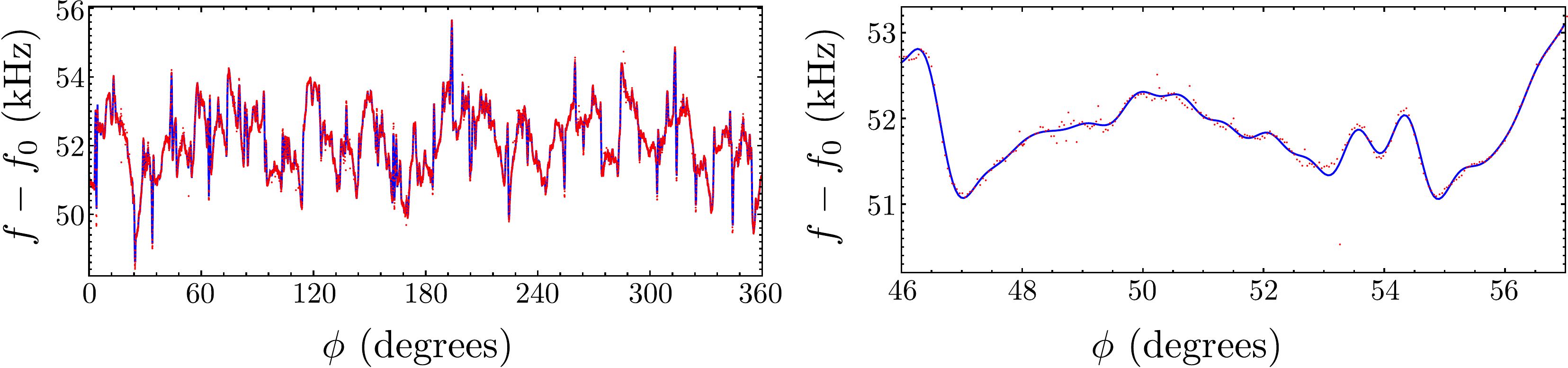}
\caption{Fourier series fit and Run~3483 data for probe~15 (upper
figures) and probe~12 (lower figures). The figures on the left show the
full range of $\phi$; the figures on the right are ``zoomed in'' to a
smaller range of $\phi$. The points (red) are the data and the curve 
(blue) is the fit.
The vertical axis is $B_z$ in units of the NMR frequency, 
offset by an amount $f_0=61.74$~MHz. For these data, 1~ppm is 
approximately $0.062$~kHz.
\label{fig:fitvsdata-3483}}
\end{figure}

Figure~\ref{fig:fitvsdata-3483} shows the Fourier series fit for
probe~15 and probe~12, along with the data as a function of $\phi$.
In contrast with the corresponding figures for Run~52, these figures
reveal the presence of large ``spikes'' in the data, for which a single
measurement is significantly displaced from the adjacent measurements.
These spikes are particularly apparent in the ``zoomed-in'' figure for
probe~12 and, in general, they are much larger in the data for probe~12
than in the data for probe~15. This suggests that these spikes may be the
reason for the large value of $\chi_q^{\rm min}(500)$ for probe~12.
In the azimuthal direction, the smallest distance scale of any
physical structure in the $g-2$ ring is about 1~cm. Some of the spikes
occur on a distance scale that is smaller than that, which suggests that
they may be artifacts of the measurement, rather than features of the
ring magnetic field. The idea that the spikes may be measurement
artifacts is bolstered by the fact that the data for Run~52, which do
not show spikes, were taken while the trolley was stationary, while the
data for Run~3483, which do show spikes, were taken while the trolley
was in motion.

\subsection{Full toroidal-harmonic fit}

Finally, we discuss the full toroidal-harmonic fit to the data, taking
$M=8$ poloidal terms for each Fourier harmonic. We have written
Mathematica code to carry out the computation of the toroidal-harmonic
coefficients. The computation of the coefficients for all of the
Fourier harmonics takes approximately 6 minutes on a dual 6-core 2.6~GHz 
Intel Xeon CPU system and consumes less than 1~GB of memory in total.

\begin{figure}[htbp]
\centering
\includegraphics[width=0.45\columnwidth]{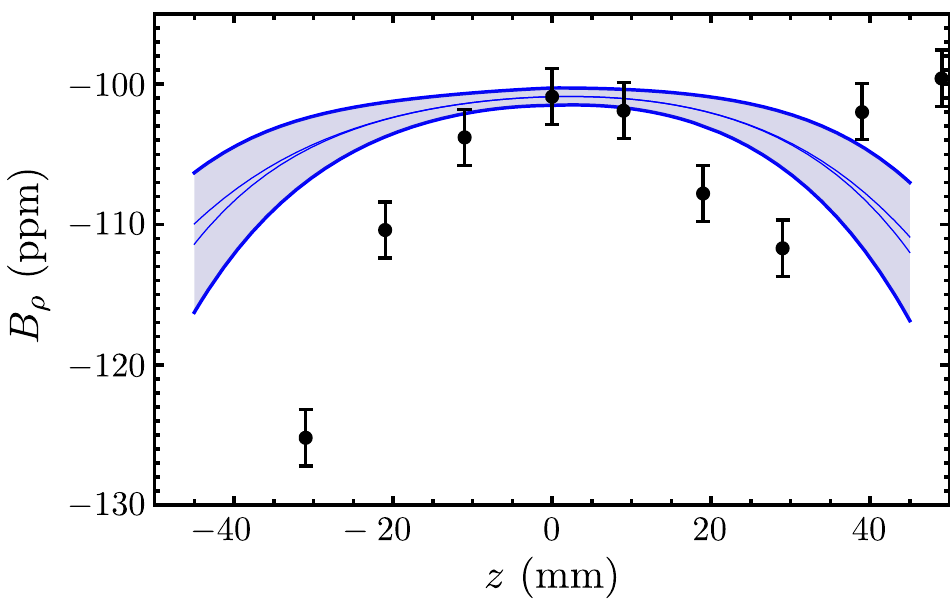}
\caption{The radial field component $B_\rho$ at $\phi=167.229^\circ$. 
Filled circles: Hall-probe data for the Run~52 field configuration.
Band: Results from the full toroidal harmonic fit to the Run~52 data,
with value at $z=0$~mm adjusted to match the Hall~probe data. The 
uncertainties that are shown in the band and for the filled circles 
are discussed in the text.
\label{fig:B-rho}}
\end{figure}

In figure~\ref{fig:B-rho}, we show Hall-probe measurements of the
radial field component $B_\rho$ and the predictions for $B_\rho$ that
come from our full toroidal-harmonic fit. (In contrast with the NMR
probes, the Hall probes can resolve the individual magnetic-field
components.) The data and the prediction are for Run~52 and correspond
to an azimuthal angle $\phi=167.229^\circ$ and to a vertical line that
passes through the center of the NMR-probe trolley.

The band in figure~\ref{fig:B-rho} shows our prediction for $B_\rho$
from the full toroidal-harmonic fits to the Run~52 NMR-probe data. In
estimating the uncertainties in the prediction, we use two measures of
the quality of our fit. One is the rms deviation of the NMR measurements
from the fit, as given in figure~\ref{fig:probes-chi-52}. The other is the
stability of the fit with respect to changes in $M$. Specifically, the
overall uncertainty in the band in figure~\ref{fig:B-rho} (thick lines)
was obtained by adding in quadrature the following quantities: (1) a
polynomial fit of order three to the rms deviations of the data from the
fit that are shown in figure~\ref{fig:probes-chi-52} and (2) the change in
the fit that is obtained by varying $M$ from 5 through 8 (thin lines in
the band). In estimating the uncertainties, we are assuming that
the physical field cannot vary rapidly in the $z$ direction in the
interval between probe measurements. This assumption is supported by the
fact that the distance between probe measurements is considerably
smaller than the distance scale of the physical structure in the $z$
direction (the diameter of the $g-2$ vacuum pipe.)

A further source of error in the full toroidal-harmonic fits arises
from the fact that the NMR trolley does not follow a perfect circle in
the azimuthal direction. Rather, it is translated by amounts that are
less than 2~mm in the radial direction and less than 1.5~mm in the
vertical direction \cite{winter}. The characteristic azimuthal angle
over which these translations occur is about $10^\circ$. The
displacements of the trolley NMR probes that arise from rotations are
much less than the displacements that arise from translations.
Consequently, the main effect of deviations of the trolley motion from
purely circular motion is to translate the field coordinates from their
true positions. Since the characteristic azimuthal angle over which these
deviations occur is much larger then either the distance between
azimuthal measurements or $2\pi/N$, the fits should be able to follow
easily these overall translations as a function of azimuthal angle.
Therefore, the trolley translations are not expected to have a
significant affect on the $\chi^2/\hbox{d.o.f}$ of the toroidal-harmonic
fits. The translations, of course, do not affect the gradients of $B_z$,
and, so, do not affect the constraints that are implied by Maxwell's
equations (\ref{maxwells-constraints}) or the results in
figure~\ref{fig:B-rho}. Examinations of the full toroidal-harmonic fits
and the raw NMR data both indicate that the $B_z$ field gradients range
from about 0.8~ppm/mm near the outer probes to about 0.11~ppm/mm near
the inner probes. Therefore, we expect the errors from overall
translations to be less than about 2~ppm near the outer probes and less
than about 0.28~ppm near the inner probes. These errors could be
corrected by translating the coordinates of the fields that are obtained
from the toroidal harmonic fits by the measured amounts of the trolley
translations.

The points in figure~\ref{fig:B-rho} show the Hall-probe
data for $B_\rho$ at $\phi=167.229^\circ$ for the Run~52 magnetic-field
configuration \cite{Hall-probe-data}. The nominal uncertainties in the 
Hall-probe measurements are 1--2~ppm. We show 2~ppm error bars in the 
figure.

We stress that, in the toroidal-harmonic fit, only $\partial
B_\rho/\partial z$ is constrained, and so only the shape of the curve
for $B_\rho$ is meaningful. As can be seen on comparison of the fit
results with the Hall-probe data, there is considerable tension
between the shapes. The data points at displacements of $-3$, $+3$,
$+4$, and $+5$ cm from the central position were taken in different
sessions than were the more central data points, and so some offsets may
have occurred. However, even the more central data points show much
larger displacements as the distance from center is varied than does the
fit. The tension between the prediction and the measurements may
indicate that there is an, as yet, unknown systematic uncertainty at
play in the Hall-probe measurements.

The main effect of the transverse fields in the $g-2$ measurement is
to induce oscillations of the beam in the transverse direction. The
Hall-probe measurements serve as an independent check of the direct
measurements of the transverse beam oscillations. The discrepancies
between the Hall-probe measurements and predictions from Maxwell's
equations that are shown in figure~\ref{fig:B-rho} are smaller than the
uncertainties in the values of the transverse fields that can be deduced
from the measurements of the beam oscillations. Hence, these 
discrepancies would not be expected to be important in the $g-2$ 
measurement. However, they may be more significant for the proposed 
measurement of the electric-dipole moment of the muon.

\section{Summary}
\label{sec:summary}

In this paper, we have presented a method for incorporating the
constraints that are imposed by Maxwell's equations into fits to the
NMR-probe data for the magnetic field of the $g-2$-experiment storage
ring. Maxwell's equations are enforced in the fits by making use of a
fitting function that consists of a truncated series of toroidal
harmonics. The toroidal harmonics are expressed as functions 
of toroidal coordinates, which are well suited to the storage-ring 
geometry.

The NMR probes measure the magnitude of the magnetic field, rather than
its individual components. It would be impractical to fit the toroidal
harmonic series for the magnitude of the magnetic field to the data
because the fitting problem is highly nonlinear and because the
transverse magnetic-field components (components that are transverse to
the average field direction) are not fully determined from the magnitude
of the field in combination with Maxwell's equations. Therefore, we have
simplified the fitting procedure by approximating the magnitude of the
field by its principal (vertical) component. Given that the transverse
components of the magnetic field in the $g-2$ storage ring are typically
of order 100 ppm, the corrections to this approximation are of order
0.01 ppm, which is comparable to the nominal error in the NMR-probe
measurements.

Even this linear fitting problem is difficult computationally.
Difficulties arise because of the large number of data points (of order
$10^5$), the large number of toroidal harmonics that is required for an
accurate fit (of order $10^4$), and the fact that the fit does not fully 
constrain the coefficients of the toroidal harmonics. A further
difficulty is that the data display fluctuations between measurements at
different azimuthal angles that are much larger than the differences
between the measurements of the various probes at a fixed azimuthal
angle. In order to ameliorate these problems, we further rearranged the
fitting procedure in to two steps. First, we fit the azimuthal
dependence of the measurements of each probe to a truncated Fourier
series. Then, we used the Fourier series coefficients for all of the
probes to determine the coefficients of the toroidal harmonics. The
underconstrained nature of the problem is confined essentially to this
second step, which involves a fitting problem whose dimensionality is 
much lower than the dimensionality of the original linear fitting problem. 
We deal with the
underconstrained nature of the problem by using the singular-value
decomposition (SVD) method, which identifies and isolates the
underconstrained modes. This modified fitting procedure results in a 
very efficient computational algorithm that is easily carried out on a 
present-day desktop computer.

We have used this method to fit some of the early NMR-probe data that
were taken by the $g-2$ collaboration. While these data are not
representative of the magnetic-field homogeneity or the
field-measurement precision that will ultimately be used by the $g-2$
collaboration to make the $g-2$ measurement, they serve to illustrate
some of the features of the fitting procedure.

Even at the level of the fits to the truncated Fourier series, it is
apparent that the deviations of the data from the fit are much larger
than the nominal precision of the NMR-probe measurements. Such fits may
be useful in diagnosing sources of uncertainties in real-world field
measurements. The full fits to the truncated toroidal-harmonic series
are in tension with Hall-probe measurements of the radial component of
the magnetic field, suggesting that there may be systematic errors in
the Hall-probe measurements that are not yet understood.

We expect that the toroidal-harmonic fits will continue to be a useful
tool in understanding the magnetic-field measurements in the $g-2$
storage ring as those measurements evolve.  The toroidal-harmonic fits
may be even more important for the planned muon
electric-dipole-moment measurement. That measurement may be more
sensitive than is the $g-2$ measurement to the transverse components of
the magnetic field. Those components are not easy to access by direct
measurements, and so the information about them that can be gleaned from
Maxwell's equations could be particularly valuable.


\begin{acknowledgments}
We thank the $g-2$ Collaboration for permission to use its preliminary
data on the $g-2$ storage-ring magnetic field. We are grateful to Joe
Grange, Ran Hong, and Peter Winter for providing us with preliminary
data from the NMR-probe measurements of the $g-2$ storage-ring magnetic
field and for many illuminating discussions. We also thank Rachel
Osofsky for providing us with preliminary data from Hall-probe
measurements of the $g-2$ storage-ring magnetic field, for several
helpful discussions, and for comments on this manuscript. The work of
G.T.B.\ is supported by the U.S.\ Department of Energy, Division of High
Energy Physics, under Contract No.\ DE-AC02-06CH11357. The work of
H.S.C.\ is supported by the Alexander von Humboldt Foundation. H.S.C.\
also acknowledges support from the Korean Research Foundation (KRF)
through the CERN-Korea fellowship program. The submitted manuscript has
been created in part by UChicago Argonne, LLC, Operator of Argonne
National Laboratory. Argonne, a U.S.\ Department of Energy Office of
Science laboratory, is operated under Contract No.\ DE-AC02-06CH11357.
The U.S.\ Government retains for itself, and others acting on its
behalf, a paid-up nonexclusive, irrevocable worldwide license in said
article to reproduce, prepare derivative works, distribute copies to the
public, and perform publicly and display publicly, by or on behalf of
the Government.

\end{acknowledgments}



\begin{thebibliography}{}
\bibitem{tech-design-report} 
E989 Collaboration, 
{\it Muon $g-2$ Technical Design Report}, 
http://lss.fnal.gov/archive/design/fermilab-design-2014-02.pdf.

\bibitem{Hall-probe-data}
R.~Osofsky, private communication.

\bibitem{hicks-1881} W.~M.~Hicks, {\it Phil.\ Trans.\ Roy.\ Soc.\ London }
{\bf 172} (1881) 609.

\bibitem{basset-1897} A.~B.~Basset, {\it Amer.\ J.\ Math. }{\bf 15} (1893) 287. 

\bibitem{milligen-fraguas-1993} B.~Ph.\ van Milligen and A.~Lopez 
Fraguas, {\it Comput.\ Phys.\ Commun. }{\bf 81} (1994) 74.

\bibitem{kuznetsov-shchepetov-1997} A.~B.~Kuznetsov and
S.~V.~Shchepetov, {\it Nucl.\ Fusion }{\bf 37} (1997) 371.

\bibitem{fischer-2012} Y.~Fischer, hal archive, hal-00679835, (2012).

\bibitem{gambini-2018} L.~Gambini, Master thesis, CERN-THESIS-2018-040, 
Bologna U., Bologna, Italy (2018).

\bibitem{abramovitz-stegun-1972} M.~Abramovitz and I.~Stegun, {\it Handbook 
of Mathematical Functions}, U.S.~Government Printing Office, Washington, 
D.C.\ (1972).

\bibitem{lawson-hanson-1995} C.~L.~Lawson and R.~J.~Hanson, {\it Solving 
Least Squares Problems}, Society for Industrial and Applied 
Mathematics, Philadelphia (1995).

\bibitem{moore-1920} E.~H.~Moore, {\it Bull.\ Amer.\ Math.\ Soc. }
{\bf 26} (1920) 394.

\bibitem{bjerhammar-1951} A.~Bjerhammar, {\it Trans.\ Roy.\ Inst.\ Tech.\ 
Stockholm }{\bf 49} (1951) 1.

\bibitem{penrose-1955} R.~Penrose, {\it Math.\ Proc.\ Cambridge\ Phil.\ Soc. }
{\bf 51} (1955) 406.

\bibitem{penrose-1956} R.~Penrose, {\it Math.\ Proc.\ Cambridge\ Phil.\ Soc. }
{\bf 52} (1956) 17.

\bibitem{grange-et-al} J.~Grange, R.~Hong, and P.~Winter, private 
communication.

\bibitem{mathematica-11} Wolfram Research Inc., {\it Mathematica},
version 11.3, Champaign, IL, U.S.A. (2018).

\bibitem{winter}  P.~Winter, private communication.                    

\end{thebibliography}
\end{document}